\documentclass[prd,twocolumn,superscriptaddress,preprintnumbers,nofootinbib,
noshowpacs,groupedaddress]{revtex4-1} 
\usepackage{epsfig}
\usepackage{color}
\usepackage{amsmath}
\usepackage{amssymb}
\usepackage{natbib}
\usepackage{enumerate}

\usepackage{graphicx,units}
\usepackage[caption=false]{subfig}
\usepackage{slashed}

\newcommand{\be}{\begin{equation}}
\newcommand{\ee}{\end{equation}}
\newcommand{\beq}{\begin{equation}}
\newcommand{\eeq}{\end{equation}}
\newcommand{\bea}{\begin{eqnarray}}
\newcommand{\eea}{\end{eqnarray}}
\newcommand{\beaa}{\begin{eqnarray*}}
\newcommand{\eeaa}{\end{eqnarray*}}
\newcommand{\ba}{\begin{array}}
\newcommand{\ea}{\end{array}}
\newcommand{\bi}{\begin{itemize}}
\newcommand{\ei}{\end{itemize}}
\newcommand{\ben}{\begin{enumerate}}
\newcommand{\een}{\end{enumerate}}

\DeclareMathOperator{\tr}{tr}
\newcommand{\ve}[1]{\mathbf{#1}}

\begin{document}
\preprint{ULB-TH/17-07}
\title{Baryogenesis from L-violating Higgs-doublet decay in the density-matrix formalism}

\author{Thomas Hambye}
\email{thambye@ulb.ac.be}
\author{Daniele Teresi}
\email{daniele.teresi@ulb.ac.be}

\affiliation{Service de Physique Th\'eorique - Universit\'e Libre de Bruxelles, Boulevard du Triomphe, CP225, 1050 Brussels, Belgium}


\begin{abstract}
We compute in the density-matrix formalism the baryon asymmetry generated by the decay of the Higgs doublet into a right-handed (RH) neutrino and a Standard Model lepton. The emphasis is put on the baryon asymmetry produced by the total lepton-number violating decay. From the derivation of the corresponding evolution equations, and from their integration, we find that this contribution is fully relevant for large parts of the parameter space. 
This confirms the results found recently in the CP-violating decay formalism with thermal corrections and shows in particular that the lepton-number violating processes are important not only for high-scale leptogenesis but also when the RH-neutrino masses are in the GeV range. For large values of the Yukawa couplings, we also find that the strong washout is generically much milder for this total lepton-number violating part than for the usual RH-neutrino oscillation flavour part. 
\end{abstract}

\maketitle

\section{Introduction}

In the framework of the type-I seesaw model of neutrino masses with right-handed (RH) neutrinos  below the electroweak scale, there exists a well-known mechanism to account for the baryon asymmetry of the Universe, through oscillations of right-handed neutrinos \cite{Akhmedov:1998qx, Asaka:2005pn, Shaposhnikov:2008pf,Canetti:2010aw,Asaka:2011wq, Drewes:2012ma, Canetti:2012kh,Shuve:2014zua, 
Canetti:2014dka,Abada:2015rta,Hernandez:2015wna, Drewes:2016gmt, Hernandez:2016kel}.
This Akhmedov-Rubakov-Smironv (ARS) scenario is based on the generation of particle-antiparticle asymmetries for the various lepton flavours. These asymmetries cancel each other in the total Standard-Model (SM) lepton-number asymmetry but, thanks to washout effects, which do not affect the different flavours in the same way, a net lepton asymmetry remains. In this framework, since the relevant processes do not involve a RH neutrino Majorana mass insertion, a lepton number can be assigned to the two helicities of the RH neutrinos. Thus, in the ARS scenario, the \emph{total} lepton number L, i.e.~the sum of the SM and the RH-neutrino ones, is conserved, but not both components separately, due to flavour effects. The SM lepton-number asymmetry component which is produced in this way before the sphalerons decouple is reprocessed in part into a baryon asymmetry, unlike the other component. The evolution of the lepton asymmetries as a function of the temperature of the thermal bath can be calculated in the density-matrix formalism, which properly takes into account the coherences between various RH neutrinos and their associated oscillations.

\begin{figure}[b!]
\includegraphics[width=0.25 \textwidth]{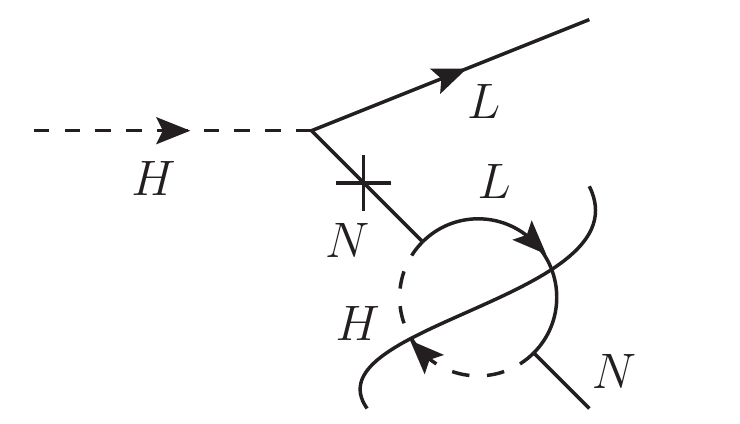}
\caption{Thermal cut in the $H \to NL$ decay, which gives rise to its purely-thermal L-violating CP-violation.\label{fig:graph}}
\end{figure}

In the different CP-violating decay formalism usually used for leptogenesis, it has been shown recently \cite{Hambye:2016sby} that the total lepton-number violating decay of the Higgs doublet into a RH neutrino and a SM lepton, i.e.~the decays which do involve a Majorana mass insertion, could also account for the baryon asymmetry. This is possible thanks to thermal effects which induce a non-zero absorptive part for the self-energy of the RH neutrino in the final state of this decay, see Fig.~\ref{fig:graph} (as this absorptive part vanishes at zero temperature). In Ref.~\cite{Hambye:2016sby} this thermal cut of the self-energy has been computed in the Kadanoff-Baym formalism. The production takes advantage of the fact that, for RH neutrino masses below the sphaleron cut, the Yukawa interactions do not thermalize the RH neutrinos so easily as for higher masses. This results in a large departure from equilibrium for the RH-neutrino number densities in the final state,  boosting the asymmetry production.

This total (SM + RH neutrino) lepton number violating scenario is in many respects different from the ARS total lepton-number conserving scenario. One difference is that it gives a non-vanishing asymmetry already for one lepton flavour.  Another is that, as a result of the fact that it involves a Majorana mass insertion, it gives rates for the relevant processes that are proportional to $m_N^2/T^2$, relatively to the total lepton-number conserving ARS piece. As a result, the asymmetry is typically produced at lower temperatures than in the ARS case, basically not long before the sphaleron decoupling. Also, since this mechanism does not require an asymmetric washout for the different flavours, its contribution in the weak-washout regime is proportional to 4 powers of the Yukawa couplings, rather than to 6 for the ARS one. 

The purpose of this paper is twofold. The first goal is to determine how this L-violating contribution can develop itself in the framework of the density-matrix formalism, rather than in the usual leptogenesis CP-violating decay formalism considered in~\cite{Hambye:2016sby}, and to compare the results obtained in both formalisms in the single lepton-flavour case, where the ARS mechanism is not operative. We will find below a good agreement between the results of both approaches showing that, also in the density-matrix formalism, the L-violating contribution is fully relevant. Moreover, as will be discussed in detail below, for large Yukawa couplings, the density-matrix formalism gives even a larger asymmetry, because the actual washout for the total lepton number is actually much smaller than the total one, considered in~\cite{Hambye:2016sby}. The second purpose is to compare, within the same formalism, which of the L-violating (LV) and L-conserving (LC) contributions dominates the production of the baryon asymmetry as a function of the parameters. In fact, evolution equations including the L-violating contribution have been derived  long ago in the density-matrix formalism \cite{Shaposhnikov:2008pf,Canetti:2012kh}, and also very recently \cite{Ghiglieri:2017gjz,Eijima:2017anv}, but this was considered as numerically irrelevant for the generation of the baryon asymmetry. This stems from the fact that L-violating processes have been thought to lead to a negligible contribution due to the $m_N^2/T^2$ suppression factor discussed above (and thus approximated as zero in the solution of the evolution equations for baryogenesis~\cite{Canetti:2012kh}) or, very recently, because the corresponding obtained rates were considered as numerically ``not significant"~\cite{Ghiglieri:2017gjz}. 
Numerically we will find below that the L-violating contribution is clearly significant, as it dominates over the flavour violating piece in large portions of the parameter space, which we determine. 

We will do all that considering only the contribution of the Higgs doublet decay/inverse decay process. Other processes, such as the top-quark and gauge scattering~\cite{Besak:2012qm,Ghiglieri:2016xye,Ghiglieri:2017gjz} and the infrared-enhanced gauge corrections to Higgs decay~\cite{Anisimov:2010gy}, are also expected to be fully relevant. But they are expected to lead to an asymmetry in a similar  way to the Higgs-doublet decay 
because they involve as a subprocess the same Higgs-doublet transition into a RH neutrino and a SM lepton.
Thus we anticipate that the relative importance of the two contributions in the various regions of parameter space is well captured by the analysis presented in this paper at the level of this decay basic building block. A fully quantitative calculation of the asymmetry produced, including these processes, is left for a further work.
All the numerical results below will be obtained considering 2 RH neutrinos, which simplifies the discussion.

\section{Density-matrix equations including lepton-number violation}

We start by deriving in this Section the density-matrix equations which allow to determine the evolution of the number density of the RH neutrinos $N_\alpha$ (which have $+$ helicity), of the conjugated states $\bar{N}_\alpha$ (which have - helicity) and of the lepton asymmetries for the various lepton flavours.  
The Lagrangian that we consider is the one of the usual type-I seesaw neutrino mass model, with nothing else,
\begin{equation}
\mathcal L_{\rm int}  \owns \ - h_{l \alpha} \bar{L}_l \widetilde{H} \mathrm{P_R} N_\alpha -\frac{1}{2} m_{N_\alpha} \bar{N}_\alpha N^c_\alpha\;+\; h.c. \;,
\end{equation}
where $\widetilde H\equiv i \sigma_2 H$ with $H=(H^+,H^0)^T$ and $L_l=(\nu_{l_L},l^-_L)^T$. In the same way as in Ref.~\cite{Hambye:2016sby}, we will make the approximation that the 4 scalar states in the doublet remain on the same footing at all relevant temperatures,  meaning in particular that they have same masses at all relevant temperatures.

The matrix number densities for the two helicities of the RH neutrinos are
\begin{align}
n^N_{\alpha \beta}(\ve k) &\equiv \frac{1}{V} \langle a^\dag_{\beta,+}(\ve k) a_{\alpha,+}(\ve k)\rangle\;, \\
\bar n^N_{\alpha \beta}(\ve k) &\equiv \frac{1}{V} \langle a^\dag_{\beta,-}(\ve k) a_{\alpha,-}(\ve k)\rangle \;, 
\end{align}
and analogously for $n^L_{l} \equiv n^L_{ll}$, $\bar n^L_{l} \equiv \bar n^L_{ll}$ (having neglected flavour coherences between the SM leptons) and $n^H \approx \bar n^{H} \approx n^H_{eq }$ (having neglected the asymmetry stored in all the SM species but the lepton doublets).

We start from the Markovian master equation~\cite{Sigl:1992fn} (see~\cite{Dev:2014laa} for a detailed derivation):
\begin{align} 
& \frac{d}{d t} n^N_{\alpha \beta}(\ve k, t) \ = \ i \,\langle [H_0^N, n^N_{\alpha \beta}(\ve k, t)]\rangle \; \notag\\ &- \; \frac{1}{2} \int_{- \infty}^{\infty}\!\!\! dt' \langle \big[ H_{\rm int}(t'), [H_{\rm int}(t), n^N_{\alpha \beta}(\ve k, t)] \big]\rangle_t   \;,
\end{align}
where $H_0^N$ is the free Hamiltonian for the RH neutrinos and where the dispersive terms, giving rise to thermal corrections to the masses, are not written explicitly.
Proceeding by a set of standard manipulations, we obtain 
\begin{align}\label{eq:transport_eq}
& \frac{d}{d t} n^N_{\alpha \beta}(\ve k) \ = \ - i \, \big[E_N, n^N(\ve k)\big]_{\alpha \beta} \notag\\
& - \frac{1}{2 E_N} \! \left( \frac{1}{2} \big\{ \Gamma^{>}(\ve k), \, n^N(\ve k)\big\} -  \frac{1}{2}\big\{ \Gamma^{<}(\ve k), \, \mathrm{I} \!-\! n^N(\ve k)\big\} \!\right)_{\alpha \beta} \!\!\!,
\end{align}
where $E_N \equiv E_N(\ve k)$. We have left the $t$-dependence implicit and the production and absorption rates $\Gamma^{\lessgtr}$ are given by 
\begin{align}\label{eq:gammas}
\Gamma_{\alpha \beta}^\lessgtr(\ve k) \ = \ &- i \tr \big\{\mathrm{P_R} u_+(\ve k) \bar u_+(\ve k) \mathrm{P_L}  \, \Sigma_{\alpha \beta}^{\lessgtr}(k) \big\} \notag \\ 
&+ i \tr \big\{\mathrm{P_R} v_+(\ve k) \bar v_+(\ve k) \mathrm{P_L}  \, \Sigma_{\beta \alpha}^{\lessgtr}(-k) \big\} \;,
\end{align}
with $k \equiv (E_N, \ve k)$. The self-energy functions $\Sigma_{\alpha \beta}^{\lessgtr}(k)$ can be expressed as
\begin{align}
- i \, \Sigma_{\alpha \beta}^{\lessgtr}(k) \ &= \ \int \! \! \frac{d^4 p}{(2 \pi)^4} \int \! \!  \frac{d^4 q}{(2 \pi)^4} \, (2 \pi)^4 \delta^{(4)}(q-k-p) \notag\\
&i S^{\lessgtr}_l(-p) \, i \Delta^{\gtrless} (-q) \, h^*_{l \alpha} h_{l \beta} \;,
\end{align}
where $i S^{\lessgtr}$ and $i \Delta^{\gtrless}$ denote, respectively, the SM-lepton and Higgs-doublet  absolutely-ordered Wightman propagators, with the Higgs-doublet one approximated to be in thermal equilibrium. Their explicit form is well-known and can be found, for instance, in~\cite{Bellac:2011kqa}. Alternatively, the transport equation~\eqref{eq:transport_eq} can be obtained directly from the field-theoretic Kadanoff-Baym formalism~\cite{Beneke:2010dz,Dev:2014wsa,Drewes:2016gmt}, once the so-called quasi-particle approximation is performed, and generalized to other processes by taking into account the appropriate self-energies. 

By assuming kinetic equilibrium for all relevant species,  we find the rate equation for the (momentum-integrated) number-density matrix $n^N_{\alpha \beta}$:
\begin{align}\label{eq:rate_eq}
& \frac{d \, n^N_{\alpha \beta}}{d t}  \, = \, - i \, \big[\mathcal E_N, n^N(\ve k)\big]_{\alpha \beta} 
\, - \, \frac{1}{2} \bigg\{\gamma^{LC} \!\! + \! \gamma^{LV}, \, \frac{n^N}{n^N_{eq}} \! - \! \mathrm{I} \, \bigg\}_{\alpha \beta} \notag\\
& + \, \frac{\delta n^L_l}{2 n^L_{eq}} \bigg( \!\big( \gamma^{LC}_{WQ,l} \!-\! \gamma^{LV}_{WQ,l}\big)  +  \frac{1}{2} \bigg\{ \gamma^{LC}_{WC,l}\! - \! \gamma^{LV}_{WC,l}, \frac{n^N}{n^N_{eq}}\bigg\}\bigg)_{\alpha \beta}   ,
\end{align}
where the thermally-averaged energy matrix is given by $\mathcal{E}_N = \text{diag}\{\mathcal{E}_{N,\alpha} \}$, with
\begin{equation}\label{eq:E_av}
\mathcal E_{N,\alpha} \ \equiv \ \frac{1}{n^N_{eq}} \, \int \!\! \frac{d^3 \ve k}{(2 \pi)^3} \, n^N_{eq}(\ve k) E_{N,\alpha}(\ve k) \;.
\end{equation}
The equilibration reaction densities are
\begin{align}
\gamma^{LC}_{\alpha \beta} \ &\equiv \ \int \! d\Pi_{\rm PS} \,  n^N_{eq}(\ve k) \big(n^L_{eq} (\ve p) + n^H_{eq}(\ve q) \big) \notag\\ &\times \,   \tr \big\{\mathrm{P_R} u_+(\ve k) \bar u_+(\ve k) \mathrm{P_L}  \, \slashed{p} \big\} \, 
h^*_{l \alpha} h_{l \beta} \;, \label{eq:gamma_LC}\\
\gamma^{LV}_{\alpha \beta} \ &\equiv \ \int \! d\Pi_{\rm PS} \,  n^N_{eq}(\ve k) \big(n^L_{eq} (\ve p) + n^H_{eq}(\ve q)\big) \notag\\ &\times \,   \tr \big\{\mathrm{P_R} v_+(\ve k) \bar v_+(\ve k) \mathrm{P_L}  \, \slashed{p} \big\} \, 
h_{l \alpha} h^*_{l \beta} \;, \label{eq:gamma_LV}
\end{align}
where the phase-space integral is
\begin{align}
\int \! d\Pi_{\rm PS} \ &\equiv \ \int\! \frac{d^3 \ve k}{(2 \pi)^3} \frac{1}{2 E_N(\ve k)} \int\! \frac{d^3 \ve p}{(2 \pi)^3} \frac{1}{2 E_L(\ve p)} \notag\\
& \times \int\! \frac{d^3 \ve q}{(2 \pi)^3} \frac{1}{2 E_H(\ve q)} \, (2 \pi)^4 \delta^{(4)}(q-p-k) \;.
\end{align}
As for the washout reaction densities, they are
\begin{align}
\gamma^{LC}_{WQ,l,\alpha \beta} \ &\equiv \ \int \! d\Pi_{\rm PS} \,  n^H_{eq}(\ve q) n^L_{eq} (\ve p) \notag\\ &\times \,   \tr \big\{\mathrm{P_R} u_+(\ve k) \bar u_+(\ve k) \mathrm{P_L}  \, \slashed{p} \big\} \, 
h^*_{l \alpha} h_{l \beta} \;, \\
\gamma^{LV}_{WQ,l,\alpha \beta} \ &\equiv \ \int \! d\Pi_{\rm PS} \,  n^H_{eq}(\ve q) n^L_{eq} (\ve p)  \notag\\ &\times \,   \tr \big\{\mathrm{P_R} v_+(\ve k) \bar v_+(\ve k) \mathrm{P_L}  \, \slashed{p} \big\} \, 
h_{l \alpha} h^*_{l \beta} \;, \\
\gamma^{LC}_{WC,l,\alpha \beta} \ &\equiv \ \int \! d\Pi_{\rm PS} \,  n^N_{eq}(\ve k) n^L_{eq} (\ve p) \notag\\ &\times \,   \tr \big\{\mathrm{P_R} u_+(\ve k) \bar u_+(\ve k) \mathrm{P_L}  \, \slashed{p} \big\} \, 
h^*_{l \alpha} h_{l \beta} \;, \\
\gamma^{LV}_{WC,l,\alpha \beta} \ &\equiv \ \int \! d\Pi_{\rm PS} \,  n^N_{eq}(\ve k) n^L_{eq} (\ve p)  \notag\\ &\times \,   \tr \big\{\mathrm{P_R} v_+(\ve k) \bar v_+(\ve k) \mathrm{P_L}  \, \slashed{p} \big\} \, 
h_{l \alpha} h^*_{l \beta} \;.
\end{align}
The ``Q'' label refers to terms which are present only due to the quantum character of the particle statistics, unlike the ``C'' labelled terms which are also present for ``classical'' Maxwell-Boltzmann statistics. 

The equation for $\bar n^N$ can be obtained easily from~\eqref{eq:transport_eq} and~\eqref{eq:gammas} by replacing the helicity-plus spinors with the minus ones in the traces, and by making use of the identities
\begin{align}
\tr \big\{\mathrm{P_R} u_+(\ve k) \bar u_+(\ve k) \mathrm{P_L}  \, \slashed{p} \big\}  &=  \tr \big\{\mathrm{P_R} v_-(\ve k) \bar v_-(\ve k) \mathrm{P_L}  \, \slashed{p} \big\} \;, \\
\tr \big\{\mathrm{P_R} v_+(\ve k) \bar v_+(\ve k) \mathrm{P_L}  \, \slashed{p} \big\}  &=  \tr \big\{\mathrm{P_R} u_-(\ve k) \bar u_-(\ve k) \mathrm{P_L}  \, \slashed{p} \big\} \;,
\end{align}
leading to
\begin{align}\label{eq:rate_eq_bar}
& \frac{d \, \bar n^N_{\alpha \beta}}{d t}  \ = \ - i \, \big[\mathcal E_N, \bar n^N(\ve k)\big]_{\alpha \beta} \notag\\
&- \, \frac{1}{2} \bigg\{\, \gamma^{LC \, *} + \gamma^{LV\, *}, \, \frac{\bar n^N}{n^N_{eq}} - \mathrm{I} \, \bigg\}_{\alpha \beta} \notag\\
& - \, \frac{\delta n^L_l}{2 n^L_{eq}} \bigg( \!\big( \gamma^{LC \,*}_{WQ,l} \! - \! \gamma^{LV\,*}_{WQ,l}\big)  + \frac{1}{2} \bigg\{ \gamma^{LC \, *}_{WC,l} - \gamma^{LV \, *}_{WC,l}, \frac{\bar n^N}{n^N_{eq}}\bigg\} \! \bigg)_{\alpha \beta} \!,
\end{align}

The equation for the asymmetry $\delta n^L_l$ can be obtained by noticing that the combinations $\sum_l \delta n^L_l + \sum_\alpha (n^N_{\alpha \alpha}-\bar n^N_{\alpha \alpha})$ and $\sum_l \delta n^L_l - \sum_\alpha (n^N_{\alpha \alpha}-\bar n^N_{\alpha \alpha})$ are not changed, respectively, by LC and LV processes. Therefore
\begin{align}\label{eq:rate_eq_asym}
\frac{d \, \delta n^L_{l}}{d t}  \ &= \ \frac{1}{n^N_{eq}} \tr \big\{\big( \gamma^{LC}_l - \gamma^{LV}_l \big) \, n^N \big\} \notag\\
&- \frac{1}{n^N_{eq}} \tr \big\{\big( \gamma^{LC \,*}_l - \gamma^{LV \,*}_l \big) \, \bar n^N \big\} \notag \\
&- \frac{\delta n^L_l}{n^L_{eq}} \tr \big\{ \gamma^{LC}_{WQ,l} + \gamma^{LV}_{WQ,l}\big\} \notag\\
& - \frac{\delta n^L_l}{2 n^L_{eq}} \frac{1}{n^N_{eq}} \tr \big\{n^N (\gamma^{LC}_{WC,l} + \gamma^{LV}_{WC,l}) \big\} \notag \\
& - \frac{\delta n^L_l}{2 n^L_{eq}} \frac{1}{n^N_{eq}} \tr \big\{ \bar n^N (\gamma^{LC \,*}_{WC,l} + \gamma^{LV \, *}_{WC,l})\big\} \;,
\end{align}
where the $l$ index is not summed. The reaction densities $\gamma^{LC}_l$ and $\gamma^{LV}_l$ are defined as in \eqref{eq:gamma_LC} and \eqref{eq:gamma_LV}, taking in them only the term with $l$ indexed Yukawa couplings.

The general structure of the evolution equations~(\ref{eq:rate_eq}), (\ref{eq:rate_eq_bar}) and (\ref{eq:rate_eq_asym}) agrees with the ones presented for the LC part in~\cite{Canetti:2012kh, Drewes:2016gmt, Hernandez:2016kel, Ghiglieri:2017gjz} and for the LV part in~\cite{Canetti:2012kh,Ghiglieri:2017gjz}, within the various approximations performed. More precisely, for the LC part these equations agree with~\cite{Hernandez:2016kel}, and also with~\cite{Canetti:2012kh, Drewes:2016gmt}, up to the fact that we do not make the extra approximation to take all washout terms as given by a same term whose form is similar to the source term, and up to the fact that \cite{Canetti:2012kh} involves an extra $1-2 n^N_{eq}(\ve k)$ factor in the equilibration rates of Eqs.~\eqref{eq:gamma_LC} and~\eqref{eq:gamma_LV} above. For the LV part we also agree with \cite{Canetti:2012kh} up to the same $1-2 n^N_{eq}(\ve k)$ factor difference. Finally with respect to the very recent Ref.~\cite{Ghiglieri:2017gjz}, the approximations made for both the LC and LV parts are not exactly the same, resulting in different equations, but the effect of these different approximations is expected to be small. 


\section{Reaction densities}  

Next in this Section we compute the ``$\gamma$" reaction densities entering in the density-matrix evolution equations above.
In the ultra-relativistic regime, i.e.~for $m_N \ll | \ve k |$, $E_N(\ve k) \simeq |\ve k| + m^2_N/(2|\ve k|)$ and from~\eqref{eq:E_av} we have
\begin{equation}
\mathcal E_{N,\alpha} \ \simeq \ \langle | \ve k | \rangle \;+\; \frac{m^2_{N,a}}{2 \, T} \bigg\langle \frac{T}{| \ve k|} \bigg \rangle \ \sim \ \frac{m^2_{N,a}}{2 \, T} \bigg\langle \frac{T}{| \ve k|} \bigg \rangle  \;,
\end{equation}
where in the last equality we have dropped the flavour independent first term because it drops in the commutators in the evolution equations. The thermal averaged $\langle T / |\ve k| \rangle$ is equal to $\simeq 0.46$ for the Fermi-Dirac statistics.
The spinor traces appearing in the lepton-number conserving and violating reaction densities are found to be
\begin{align}
\tr \big\{\mathrm{P_R} u_+(\ve k) \bar u_+(\ve k) \mathrm{P_L}  \, \slashed{p} \big\} \ &=\ (E_N + | \ve k|)(p^0 - |\ve p| \cos \theta) \notag\\
&\simeq \  2 | \ve k| (p^0 - |\ve p| \cos \theta) \;, \\
\tr \big\{\mathrm{P_R} v_+(\ve k) \bar v_+(\ve k) \mathrm{P_L}  \, \slashed{p} \big\} \ &=\ (E_N - | \ve k|)(p^0 + |\ve p| \cos \theta) \notag\\
&\simeq \  \frac{m_N^2}{2 | \ve k|} (p^0 + |\ve p| \cos \theta) \;,
\end{align}
where $\theta$ is the angle between $\ve k$ and $\ve p$. One recognizes that for negative helicity, i.e.~for LV transitions, a $m_N^2$ factor appears, as it should be, since LV requires a Majorana mass insertion. 
For the washout reaction densities, denoting by a tilde their value once the Yukawa couplings have been factorized out, e.g. $\gamma^{LC}_{l, \alpha \beta} \equiv \widetilde{\gamma}^{LC} h_{l \alpha}^* h_{l \beta}$, we find 
\begin{align}
&\widetilde{\gamma}^{LC} \ = \ \frac{g_{iso}}{4 \pi^2} \int_0^\infty \!\!\! dk \, \frac{k}{e^{k/T}+1}
 \frac{M_H^2-M^2_L}{8 \pi k}  \notag\\ &\times \int_{E^*}^\infty \!\!\! dE \, \bigg(\frac{1}{e^\frac{E}{T}+1} + \frac{1}{e^\frac{E+k}{T}-1} \bigg) \notag \\
 &\simeq 3.26 \times 10^{-4} \, T^4 \ \equiv \alpha^{LC} \, T^4 \;, 
 \label{LCrate}\\[9pt]
 &\widetilde{\gamma}^{LV} \ = \ \frac{g_{iso}}{4 \pi^2} \int_0^\infty \!\!\! dk \, \frac{k}{e^{k/T}+1}
 \frac{m_N^2}{32 \pi k^3}  \notag\\ &\times \int_{E^*}^\infty \!\!\! dE \, (4 E k +M^2_L - M^2_H) \bigg(\frac{1}{e^\frac{E}{T}+1} + \frac{1}{e^\frac{E+k}{T}-1} \bigg) \notag \\
 &\simeq 3.35 \times 10^{-3} \, m^2_N \, T^2 \ \equiv \ \alpha^{LV} m_N^2 \, T^2\;,
  \label{LVrate} \displaybreak[0]\\[9pt]
 &\widetilde{\gamma}_{WQ}^{LC} \ = \ \frac{g_{iso}}{4 \pi^2} \int_0^\infty \!\!\! dk \, k \,
 \frac{M_H^2-M^2_L}{8 \pi k}  \notag\\ &\times \int_{E^*}^\infty \!\!\! dE \,\frac{1}{e^\frac{E}{T}+1} \frac{1}{e^\frac{E+k}{T}-1} \notag \\
 &\simeq 1.05 \times 10^{-4} \, T^4 \ \equiv \alpha_W^{LC} \, T^4 \;,
  \label{WQLCrate} \\[9pt]
 &\widetilde{\gamma}_{WQ}^{LV} \ = \ \frac{g_{iso}}{4 \pi^2} \int_0^\infty \!\!\! dk \, k \,
 \frac{m_N^2}{32 \pi k^3}  \notag\\ &\times \int_{E^*}^\infty \!\!\! dE \, (4 E k +M^2_L - M^2_H) \bigg(\frac{1}{e^\frac{E}{T}+1} + \frac{1}{e^\frac{E+k}{T}-1} \bigg) \notag \\
 &\simeq 5.49 \times 10^{-4} \, m^2_N \, T^2 \;,
  \label{WQLVrate} \displaybreak[0]\\[9pt]
&\widetilde{\gamma}_{WC}^{LC} \ = \ \frac{g_{iso}}{4 \pi^2} \int_0^\infty \!\!\! dk \, \frac{k}{e^{k/T}+1} \,
 \frac{M_H^2-M^2_L}{8 \pi k} \int_{E^*}^\infty \!\!\! dE \,\frac{1}{e^\frac{E}{T}+1}  \notag \\
 &\simeq 1.86 \times 10^{-4} \, T^4  \;,
  \label{WCLCrate} \\[9pt]
 &\widetilde{\gamma}_{WC}^{LV} \ = \ \frac{g_{iso}}{4 \pi^2} \int_0^\infty \!\!\! dk \, \frac{k}{e^{k/T}+1} \,
 \frac{m_N^2}{32 \pi k^3}  \notag\\ &\times \int_{E^*}^\infty \!\!\! dE \, (4 E k +M^2_L - M^2_H) \, \frac{1}{e^\frac{E}{T}+1} \notag \\
 &\simeq 1.79 \times 10^{-3} \, m^2_N \, T^2 \;,
  \label{WCLVrate}
\end{align}
where (neglecting $m_N$ compared to the other masses)
\begin{equation}
E^* \ \equiv \ \frac{k \,M^2_L}{M_H^2 - M^2_H} \,+\,  \frac{M_H^2-M_L^2}{4 \,k} \;,
\end{equation}
and we have included $g_{iso} = 2$ to account for the two isospins of the processes (e.g.~$H^0 \to N \bar \nu$ and $H^+ \to N e^+$). For the thermal masses of $H$ and $L$ we have used the high-temperature asymptotic masses (i.e.~the ones in the large 3-momentum limit)
\begin{align}
M_H^2 \ &= \ \frac{T^2}{16} (3 g^2 + g'^2 + 4 h_t^2 + 8 \lambda) \;,\\
M_L^2 \ &= \ \frac{T^2}{16} (3 g^2 + g'^2) \;.
\end{align}

\section{Analytic solutions in the weak-washout and many-oscillations regime}\label{sec:analytic}

Before presenting the numerical results obtained by integrating the density-matrix evolution equations, it is instructive to compare the LC and LV analytic solutions which can be obtained in the weak-washout regime for the case of many oscillations taking place, i.e.~when oscillations start long before sphaleron decoupling.
The derivation of these analytic solutions is given in Appendix~\ref{app:analytic}. They are determined at first non-vanishing order in powers of the Yukawa couplings. We give them for the lepton asymmetry yield for each flavour $l$, defined as the lepton asymmetry obtained for this flavour over the entropy density, i.e.~$Y_l\equiv n_l^L/s$. Notice that in the weak-washout regime considered in this section the total asymmetry is simply given by the sum of the ARS and the LV H-decay one.
For
the LC contribution, summing over the lepton flavour, $Y \equiv \sum_l Y_l$, at the sphaleron decoupling temperature, $T=T_s\simeq 131.7$~GeV where the baryon asymmetry freezes, we get the following total lepton asymmetry, which involves 6 powers of the Yukawa couplings,
\begin{align}
Y_{LC} \ &\simeq  - \, 18.5 \times (\alpha^{LC})^2 \, \alpha^{LC}_W \, \frac{M_0^{7/3}}{T_c (\Delta m^2_N)^{2/3}} \,\notag\\
&\times \, (h^\dag h)_{11} (h^\dag h)_{22} \sum_l \delta_l^{LC} (h h^\dag)_{ll} \;,
\label{LFanalytsol}
\end{align}
with  
\begin{equation}
\delta^{LC}_l \ = \ \frac{\mathrm{Im}\big[h_{l1}^* h_{l2} (h^\dag h)_{21}\big]}{(h^\dag h)_{11} (h^\dag h)_{22}} \;,
\end{equation}
and $M_0 \simeq M_{PL}/(1.66 \, g_*^{1/2}) \simeq \unit[7 \times 10^{17}]{GeV}$. This result involves three rates whose value are given in the previous Section.
For the LV contribution, instead, one gets an asymmetry which involves 4 powers of the Yukawa couplings and 2 rates
\begin{align}
Y_{LV} \ &\simeq \ 7.9 \times \alpha^{LC} \, \alpha^{LV} \, \frac{M_0}{T_c} \,\frac{m_N^2}{\Delta m_N^2} \, (h^\dag h)_{11} (h^\dag h)_{22} \, \delta^{LV} \;,
\label{LVanalytsol}
\end{align}
with  
\begin{equation}
\delta^{LV}_l \ = \ \frac{\mathrm{Im}\big[h_{l1}^* h_{l2} (h^\dag h)_{12}\big]}{(h^\dag h)_{11} (h^\dag h)_{22}}\;,
\end{equation}
and $\delta^{LV} \equiv \sum_l \delta^{LV}_l$.

Comparing Eqs.~\eqref{LVanalytsol} and~\eqref{LFanalytsol} one observes a series of differences.
First of all, both contributions involve different combinations of the Yukawa coupling, $\delta^{LV}$ versus $\sum_l \delta_l^{LC} (h h^\dag)_{ll}$ for the LV and LC contributions respectively. As already mentioned in the Introduction, the LC combination vanishes in the case of a single flavour, unlike the LV one. Generically, the combination of Yukawa couplings in the LV case is much larger because it involves 2 Yukawa couplings less.  Moreover, unless one chooses precisely specific phases in the Yukawa entries, 
the ratio of $\sum_l \delta_l^{LC} (h h^\dag)_{ll}$ to $\delta^{LV}$ turns out to be smaller than the naive estimate $m_\nu m_N/v^2$ by one order of magnitude or more.
Next, these 2 contributions display different powers of the Majorana mass scale, of the mass splitting and of the  effective Planck scale, $M_0 m_N^2/\Delta m_N^2$ versus $M_0^{7/3}/ (\Delta m_N^2)^{2/3}$. As discussed above, the $m_N^2$ factor results from the Majorana-mass insertion necessary for the LV contribution. The different powers of the mass splitting result from the different $T$-dependence of the LC and LV rates, and in particular from the relative size of the commutator term and the ``$\gamma$'' rates as functions of the temperature.  As for the different powers in the effective Planck mass $M_0$, they result from the integration over $t$ of the different temperature dependence of the contributions. Such a scale ratio is clearly in favour of the LC contribution, especially for small $m_N$ and large mass splittings. For instance, with $m_N=10$~GeV and $\Delta m_N/m_N=10^{-10}$ or with $m_N=1$~GeV and $\Delta m_N/m_N=10^{-5}$ one gets a relative factor equal to 
$5.9 \times 10^{-20}$
and $5.9 \times 10^{-23}$ respectively.
Finally, beside a different numerical factor in favour of the LC contribution, -18.5 versus 7.9, the rate factors are very different, $\alpha^{LC} \, \alpha^{LV} $ versus $ (\alpha^{LC})^2 \, \alpha^{LC}_W$. As said above, the LV piece involves 2 rates whereas the LC piece involves 3 of them, as a result of the fact that it requires the asymmetric washout to play a role. Hence, the rate factor is much larger for the LV contribution. Numerically, from Eqs.~(\ref{LCrate})-(\ref{WQLCrate}), the rate ratio of both contribution is
$\alpha^{LC} \, \alpha^{LV}/  [(\alpha^{LC})^2 \, \alpha^{LC}_W]=0.97\times 10^5$.

All in all the various factors compete in such a way that, depending on the input parameters, one contribution or the other can be dominant in large portions of the parameter space. Such a comparison will be made in detail in Section~\ref{sec:comparison} by solving the transport equations numerically. In the meanwhile one could just give two examples at the level of the approximate solutions, valid in the small-washout regime. For $m_N=10$~GeV and $\Delta m_N/m_N=10^{-10}$ or with $m_N=1$~GeV and $\Delta m_N/m_N=10^{-5}$,
the ratio of the LV to LF asymmetries obtained in Eqs.~(\ref{LVanalytsol}) and (\ref{LFanalytsol}) is $c \cdot (\delta^{LV}/\sum_l \delta_l^{LC} (h h^\dag)_{ll})$ with $c=-2.4 \times 10^{-15}$ and $c=-2.4 \times 10^{-18}$, respectively. For Yukawa couplings of order $(\sqrt{m_\nu m_N/v^2})^{1/2}$, i.e.~$(\delta^{LV}/\sum_l \delta_l^{LC} (h h^\dag)_{ll}) \gtrsim 10^{15}$, $10^{16}$, this gives the LV and LC contributions dominant respectively. 

\section{Phenomenology}

\subsection{Results with a single lepton flavour: comparison of density-matrix and CP-violating decay formalisms}\label{sec:oneleptonflavour}

To compare the results obtained above for the LV contribution in the density-matrix formalism with the ones obtained in the CP-violating decay formalism~\cite{Hambye:2016sby}, it is convenient to consider only one lepton flavour and 2 RH neutrinos. Considering only one flavour makes sense because in this case the LC ARS contribution identically vanishes and does not interplay  with the LV contribution in a non-linear way (as it does, instead, in the strong washout regime with several flavours). 

With one lepton flavour and 2 RH neutrinos there is only a very limited number of parameters, the RH neutrino scale $m_N$, the mass splitting between the two RH neutrinos, $\Delta m_N$, the Yukawa coupling of $N_1$ which can be taken to be real, $h_1$, and the Yukawa coupling of $N_2$ which  then has a phase, $h_2=|h_2|e^{i\theta}$. The last 3 parameters can be traded for the effective mass $\tilde{m}\equiv v^2 h_1^2/m_N$, which parametrizes the speed of the decay of $N_1$, for the ratio of decay widths $\Gamma_{22}/\Gamma_{11}$ and for the CP-violating parameter $I_1=\delta^{LV}$.

\begin{figure}
\includegraphics[width = 0.8 \columnwidth]{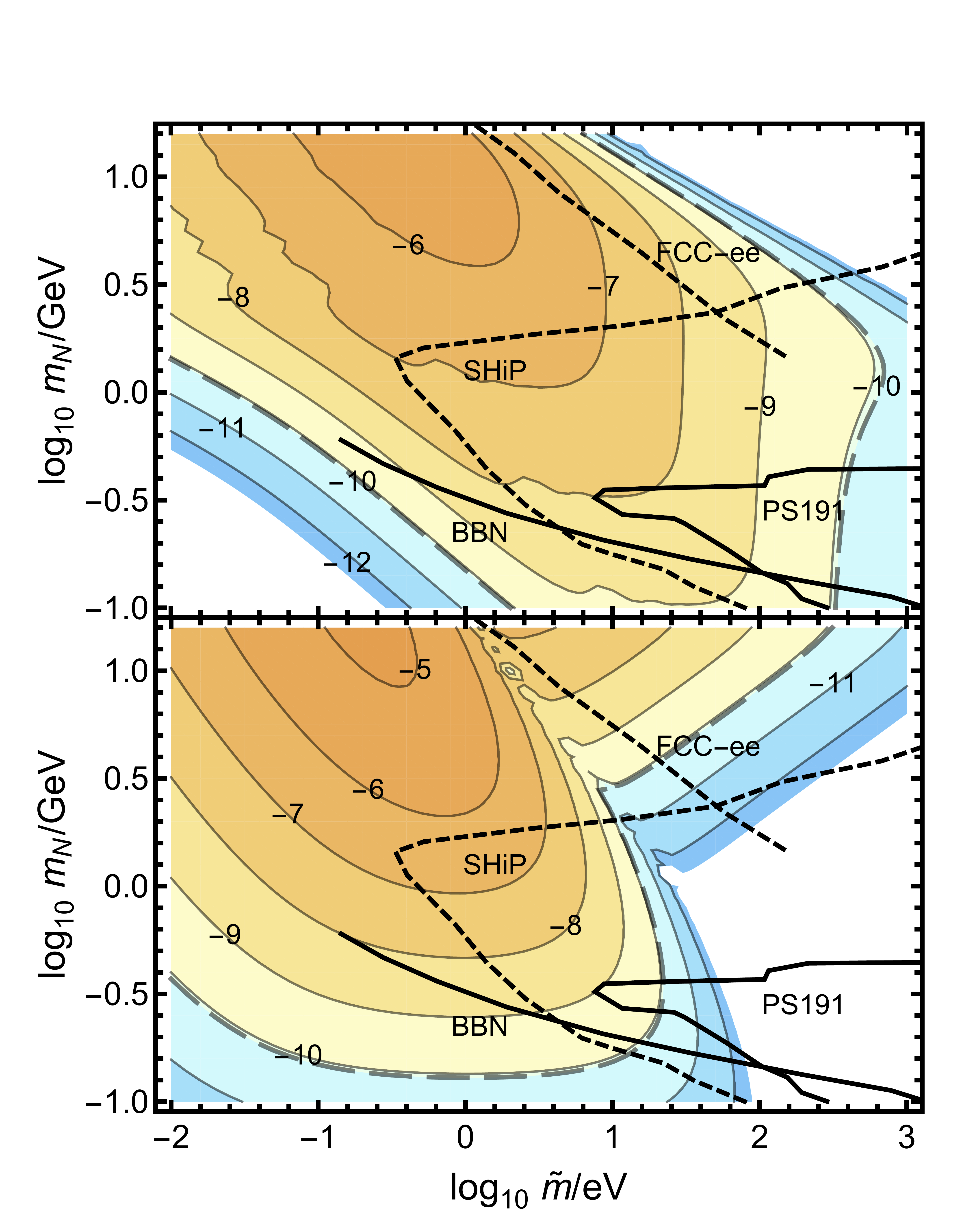}
\caption{Baryon asymmetry $n_B/s$ obtained a a function of $m_N$ and $\tilde{m}$ for $\Delta m_N/m_N=10^{-11}$, $I_1=1$  and 
$\Gamma_{11}/\Gamma_{22}=m_{sol}/m_{atm}$. Upper panel: from denstity matrix formalism, Eqs.~(\ref{eq:rate_eq}), (\ref{eq:rate_eq_bar}) and (\ref{eq:rate_eq_asym}). Lower panel: from CP-violating decay formalism, as in Ref.~\cite{Hambye:2016sby}. The successful regions are the ones inside the dashed line which corresponds to the observed value $n_B/s=0.86 \times 10^{-10}$.}
\label{fig:2N_1L_dmn_11}
\end{figure}

\begin{figure}
\includegraphics[width = 0.8 \columnwidth]{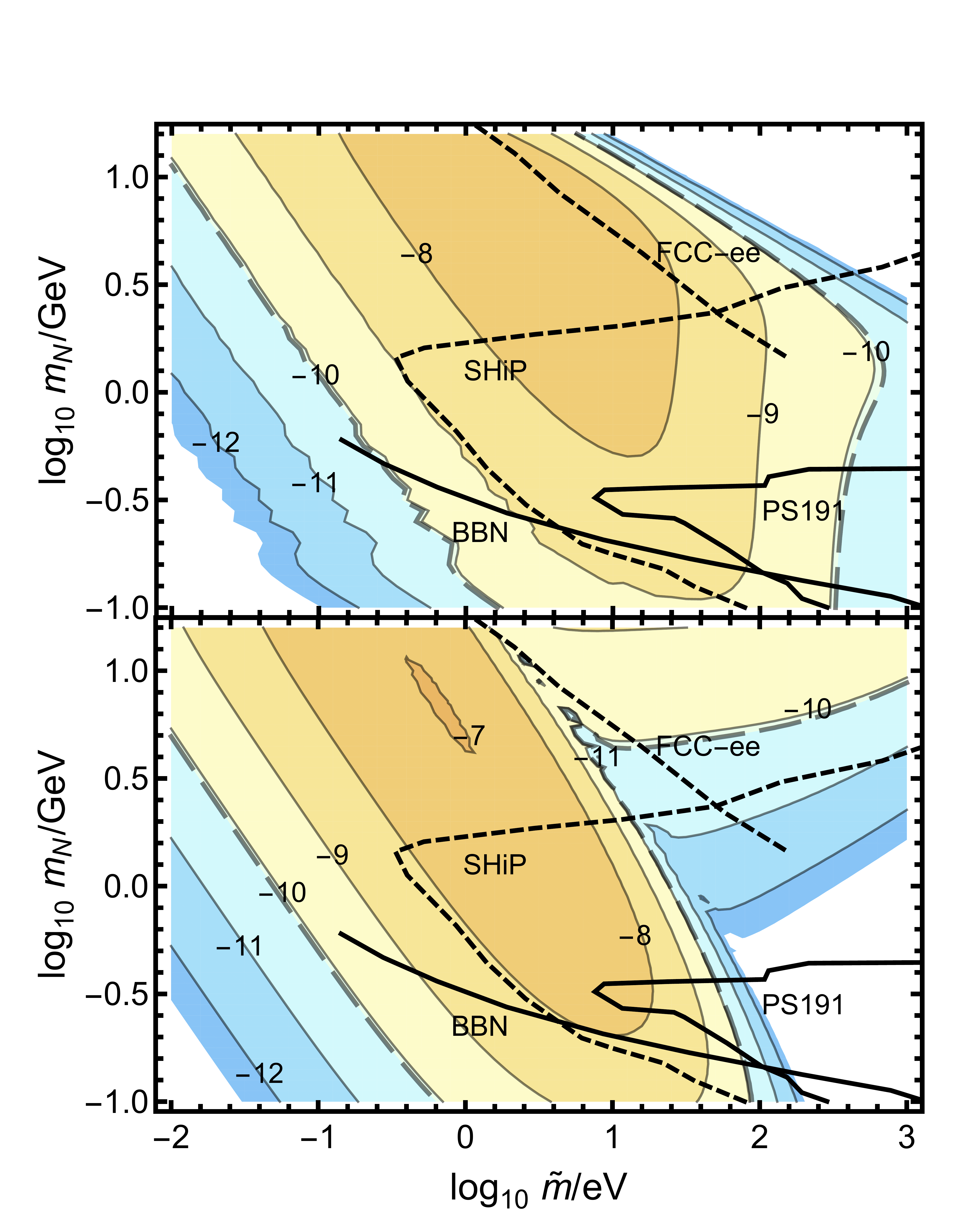}
\caption{Same as Fig.~\ref{fig:2N_1L_dmn_11}, for $\Delta m_N/m_N=10^{-8.5}$.}
\label{fig:2N_1L_dmn_8_5}
\end{figure}

Starting from a situation where there is no RH neutrinos at $T >>> T_s$, in Figs.~\ref{fig:2N_1L_dmn_11} and \ref{fig:2N_1L_dmn_8_5} we plot the results for the baryon asymmetry obtained by integrating numerically Eqs.~(\ref{eq:rate_eq}), (\ref{eq:rate_eq_bar}) and (\ref{eq:rate_eq_asym}). This is shown as a function of $m_N$ and $\tilde{m}$ for the same value of the other parameters as considered in \cite{Hambye:2016sby}.
Also shown on these figures are the results obtained from the CP-violating decay formalism in \cite{Hambye:2016sby}.
These figures nicely show a same qualitative pattern. Quantitatively, for not too small $m_N$,  $\Delta m_N/m_N$ and not too large Yukawa couplings, the asymmetries obtained are in general of the same order of magnitude. For small $m_N$ and $\Delta m_N/m_N$ the difference is due to the simplified treatment of RH-neutrino thermal masses in the density-matrix equations above, which would necessitate a $T$-dependent diagonalization of the $N$ mass matrix, not taken into account here for simplicity.
For large $\tilde{m}$ the fact that the produced asymmetry obtained can be much larger in the density matrix formalism is not due to a real discrepancy between both methods. This difference stems from the fact that in \cite{Hambye:2016sby} the washout suppression has been overestimated.
As showed above, in the density-matrix formalism one gets 2 different types of washout. First there is the usual ARS washout term $\gamma^{LC}_W\propto T^4$, Eq.~(\ref{WQLCrate}), which violates SM lepton number but does not violate total lepton number (as it does nothing more than transferring the asymmetry between the SM-lepton and RH-neutrino sectors). This piece washes out the ARS LC asymmetry produced, but does not wash out the total lepton-number asymmetry produced.  Second, there is the $\gamma^{LV}_W\propto m_N^2 T^2$ washout part, Eq.~(\ref{WQLVrate}), which violates total lepton number but is not as effective, because suppressed by a factor of $m_N^2/T^2$ with respect to the other washout.
In the CP-violating decay formalism of \cite{Hambye:2016sby},  the fact that a lepton-number asymmetry can be stored in the RH-neutrino sector is not taken into account (but could, in principle, in a more involved decay formalism). Thus, the part of the washout $\propto T^4$ incorrectly washes out the final lepton lepton asymmetry too, resulting in an overestimation of the washout suppression. In the density-matrix formalism this is instead correctly taken into account.


\subsection{Results with 3 lepton flavours: comparison of the LC results with the total LC + LV results}\label{sec:comparison}

\begin{figure*}[p]
\includegraphics[width=0.33\textwidth]{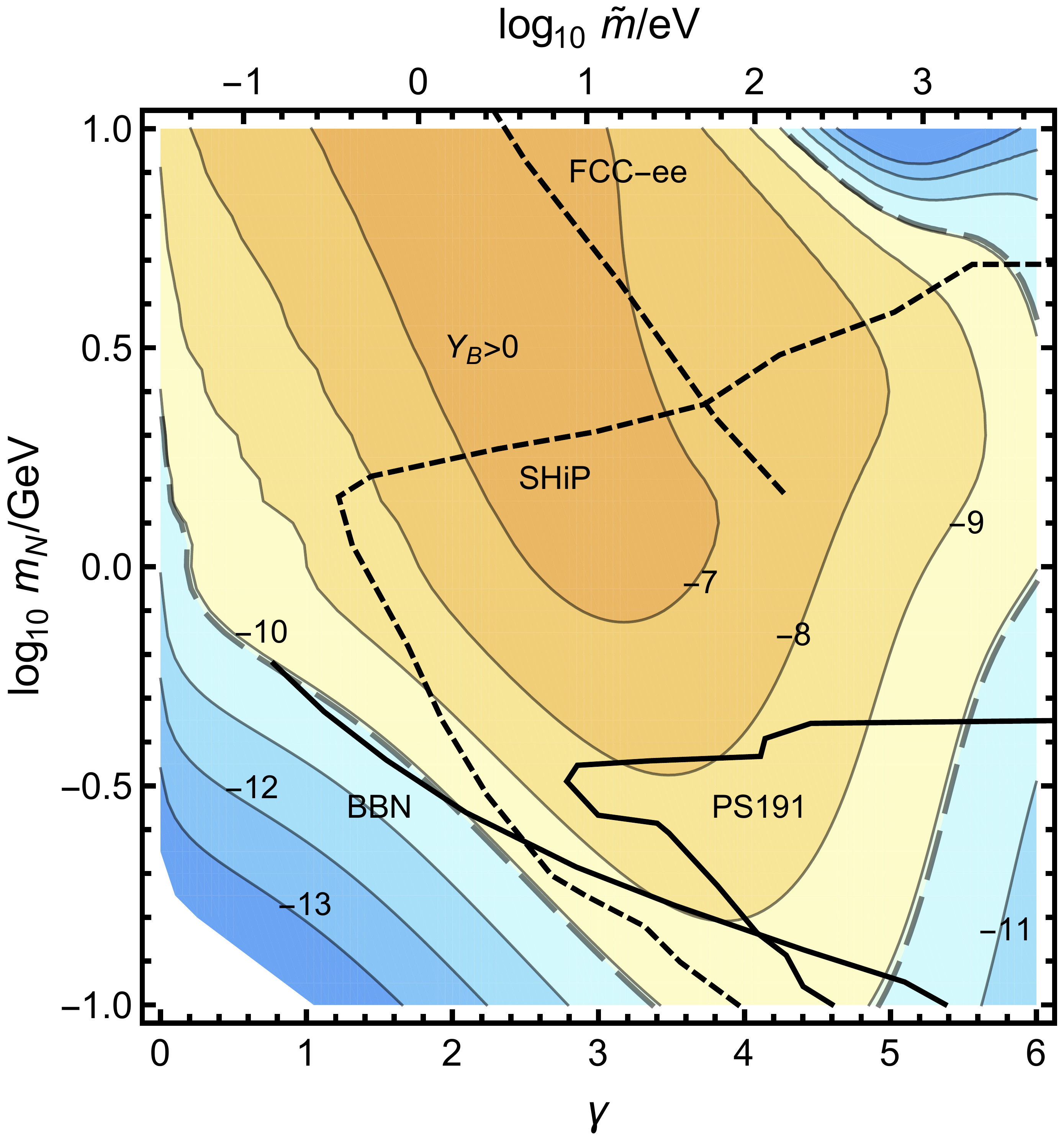}\qquad\qquad\qquad
\includegraphics[width=0.33\textwidth]{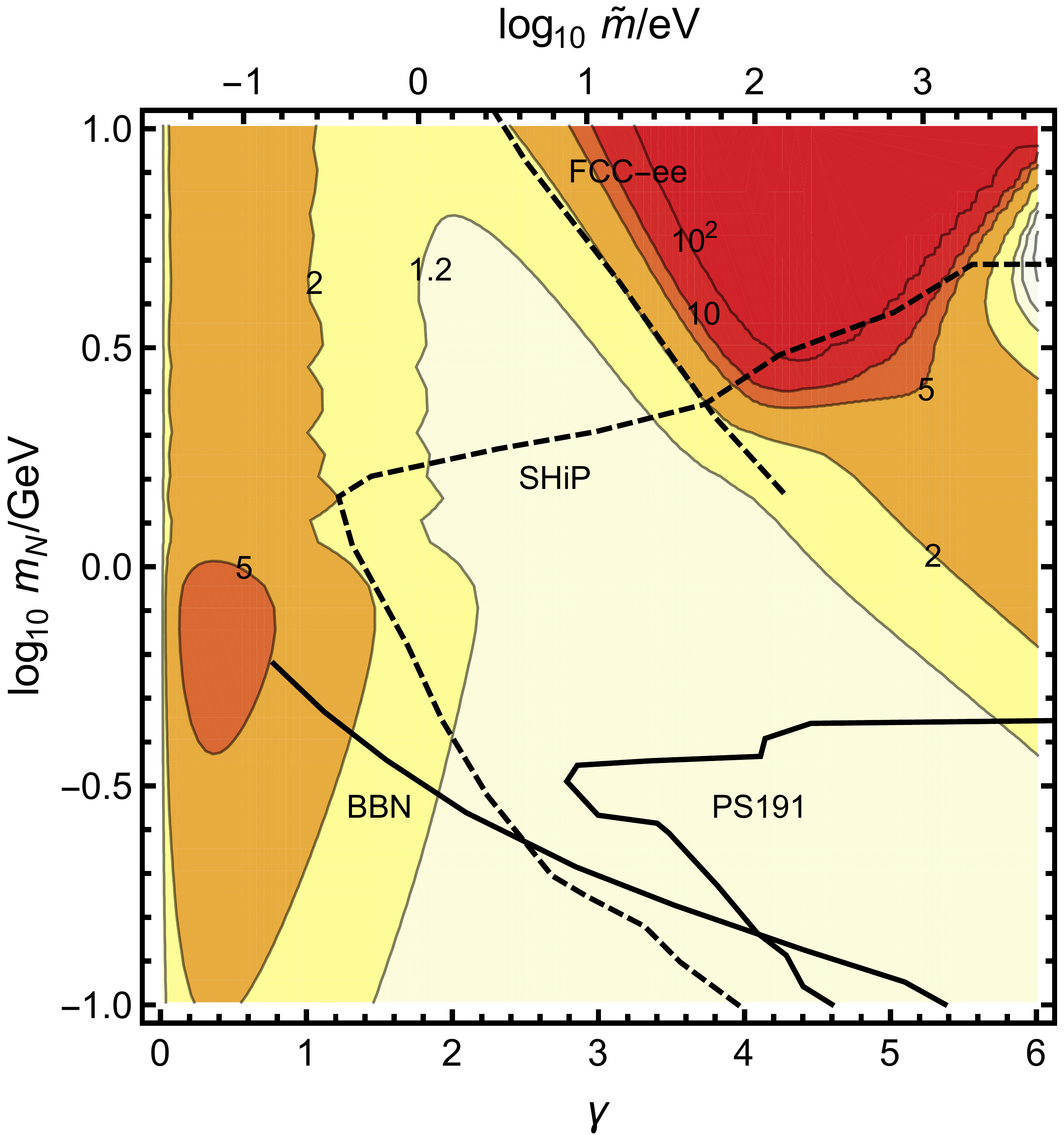}
\caption{Results for $\Delta m_N/m_N = 10^{-10}$, $\alpha_2=0$\label{fig:10_A}. In the left panel, we plot the logarithm base 10 of the $n_B/s$ asymmetry obtained. The observed value, $n_B/s \simeq 0.86 \times 10^{-10}$ is denoted by the dashed line. In the right panel, we plot the ratio of the full LC + LV result to the LC ARS one.}
\end{figure*}
\begin{figure*}[p]
\includegraphics[width=0.33\textwidth]{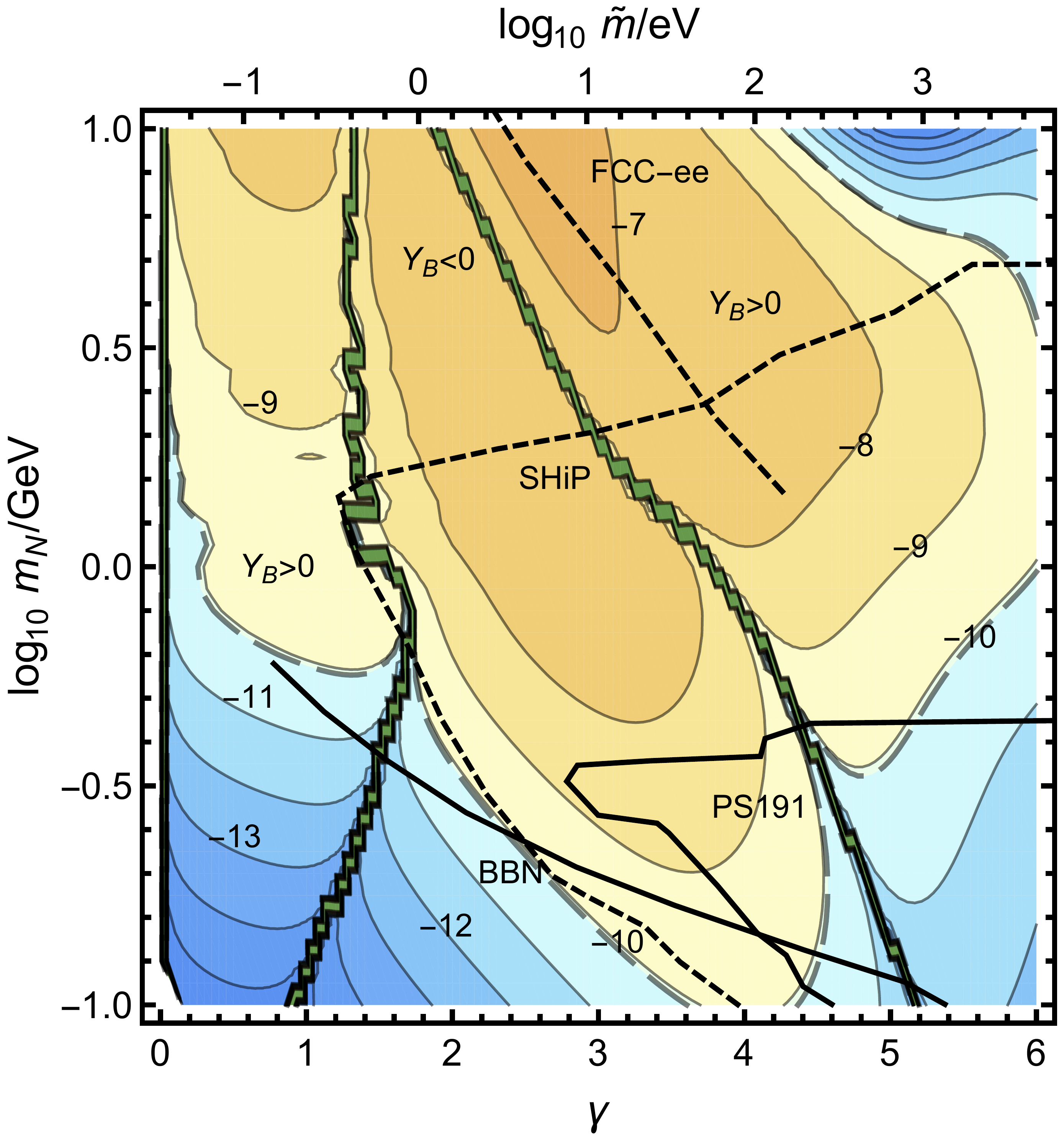}\qquad\qquad\qquad
\includegraphics[width=0.33\textwidth]{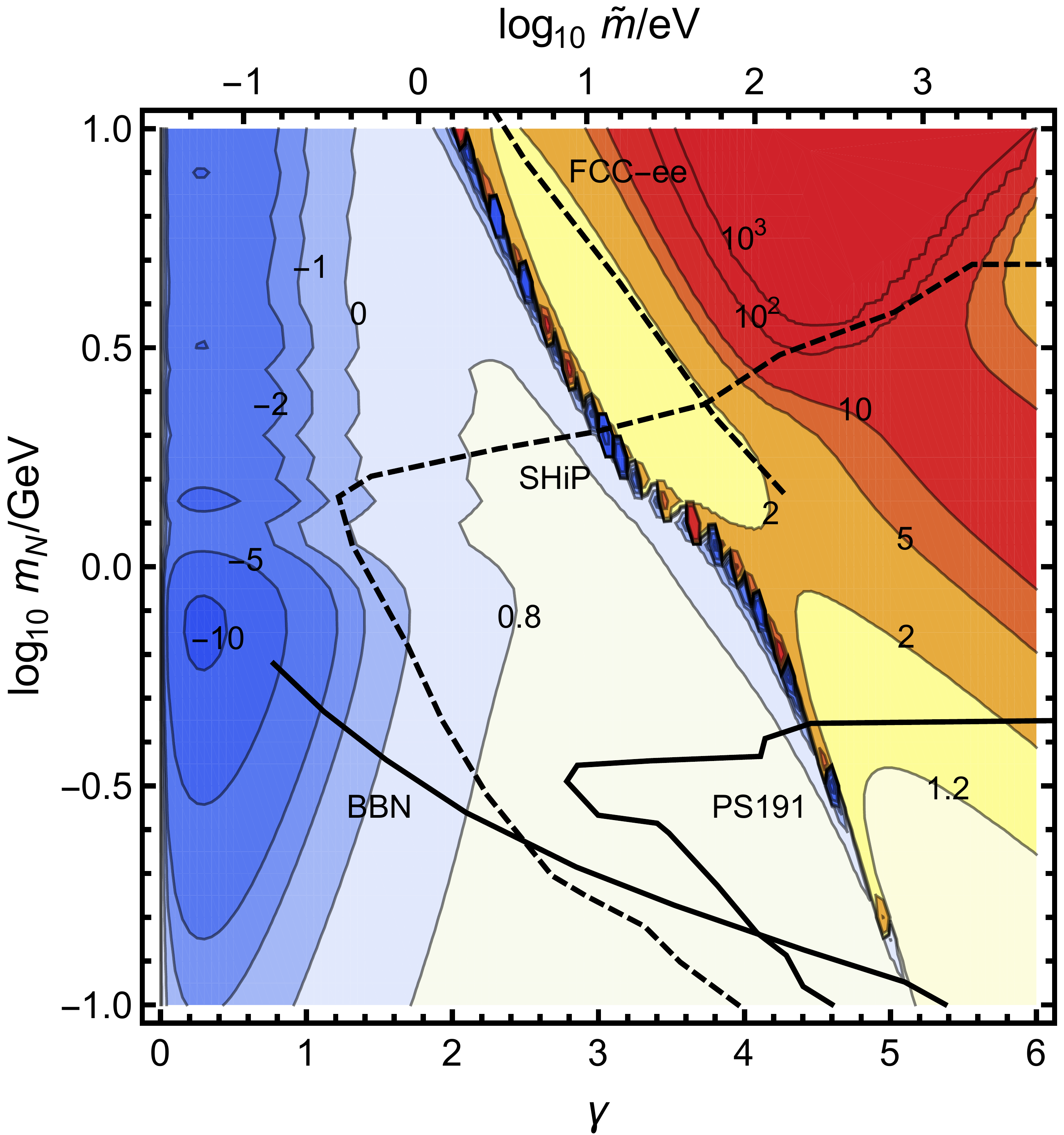}
\caption{Results for $\Delta m_N/m_N = 10^{-10}$, $\alpha_2=\pi/2$. The dark green band signals a change of sign of the final asymmetry obtained.}
\end{figure*}
\begin{figure*}[p]
\includegraphics[width=0.33\textwidth]{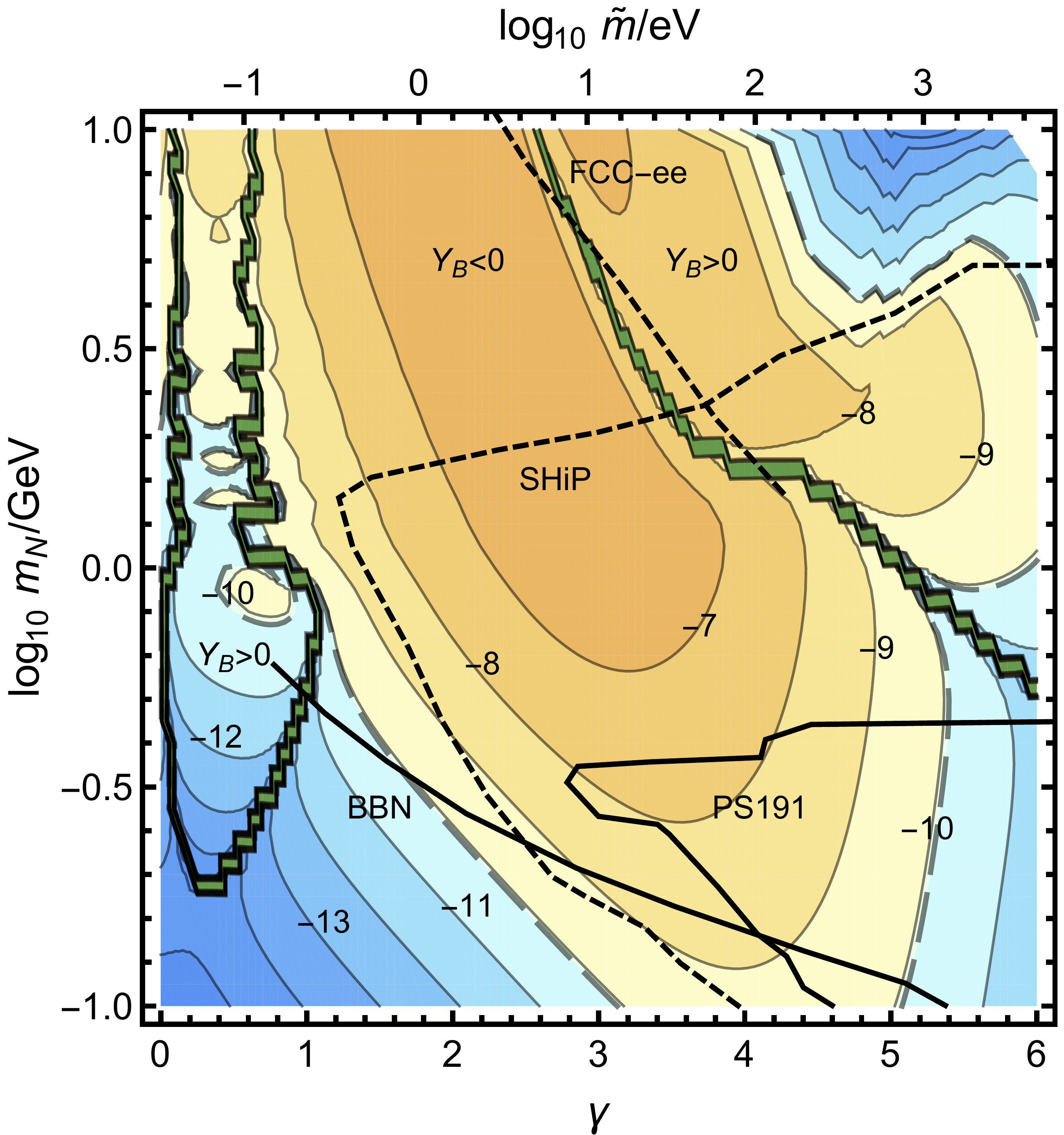}\qquad\qquad\qquad
\includegraphics[width=0.33\textwidth]{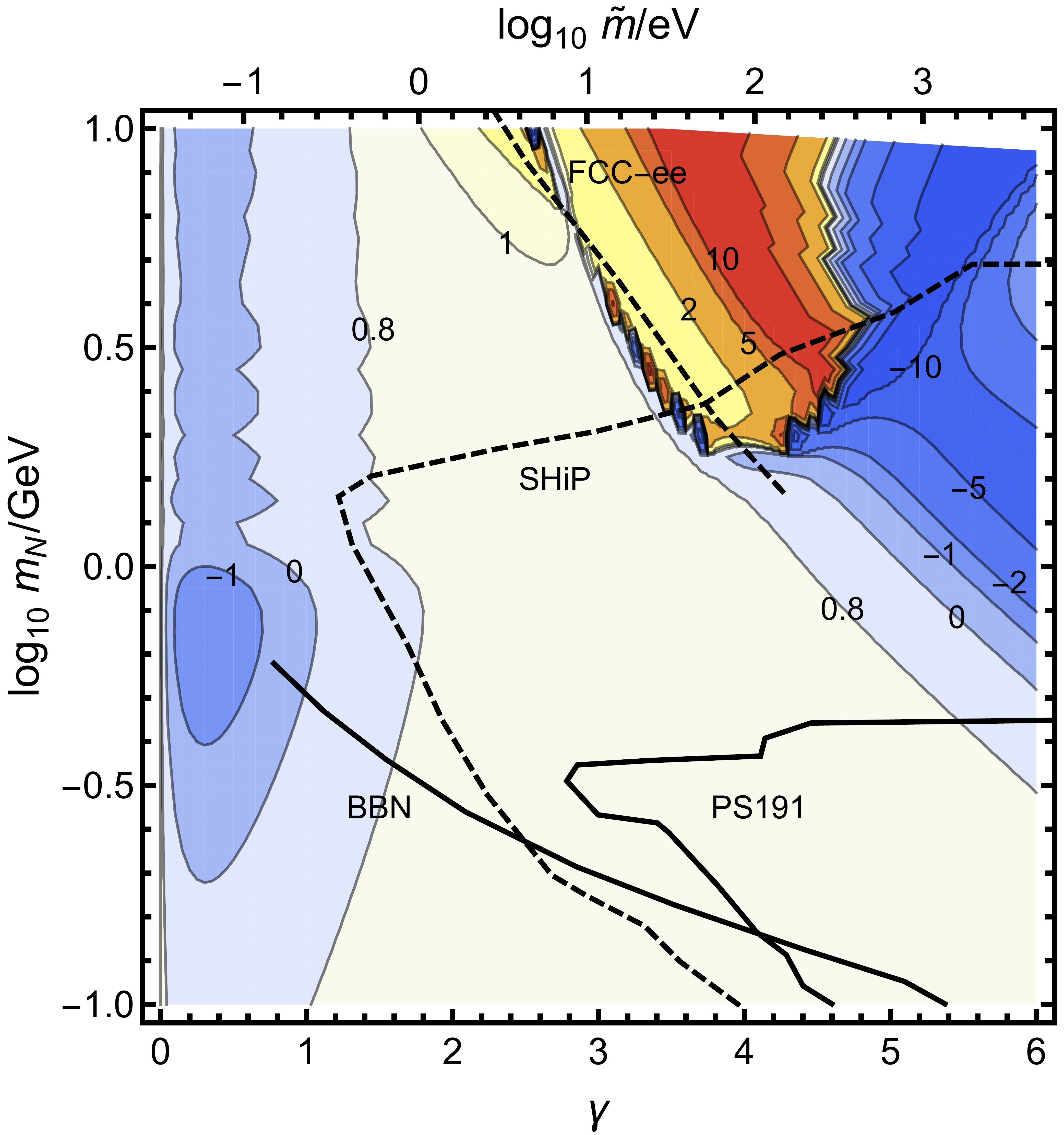}
\caption{Results for $\Delta m_N/m_N = 10^{-10}$, $\alpha_2=\pi$.}
\end{figure*}

\begin{figure*}[p]
\includegraphics[width=0.33\textwidth]{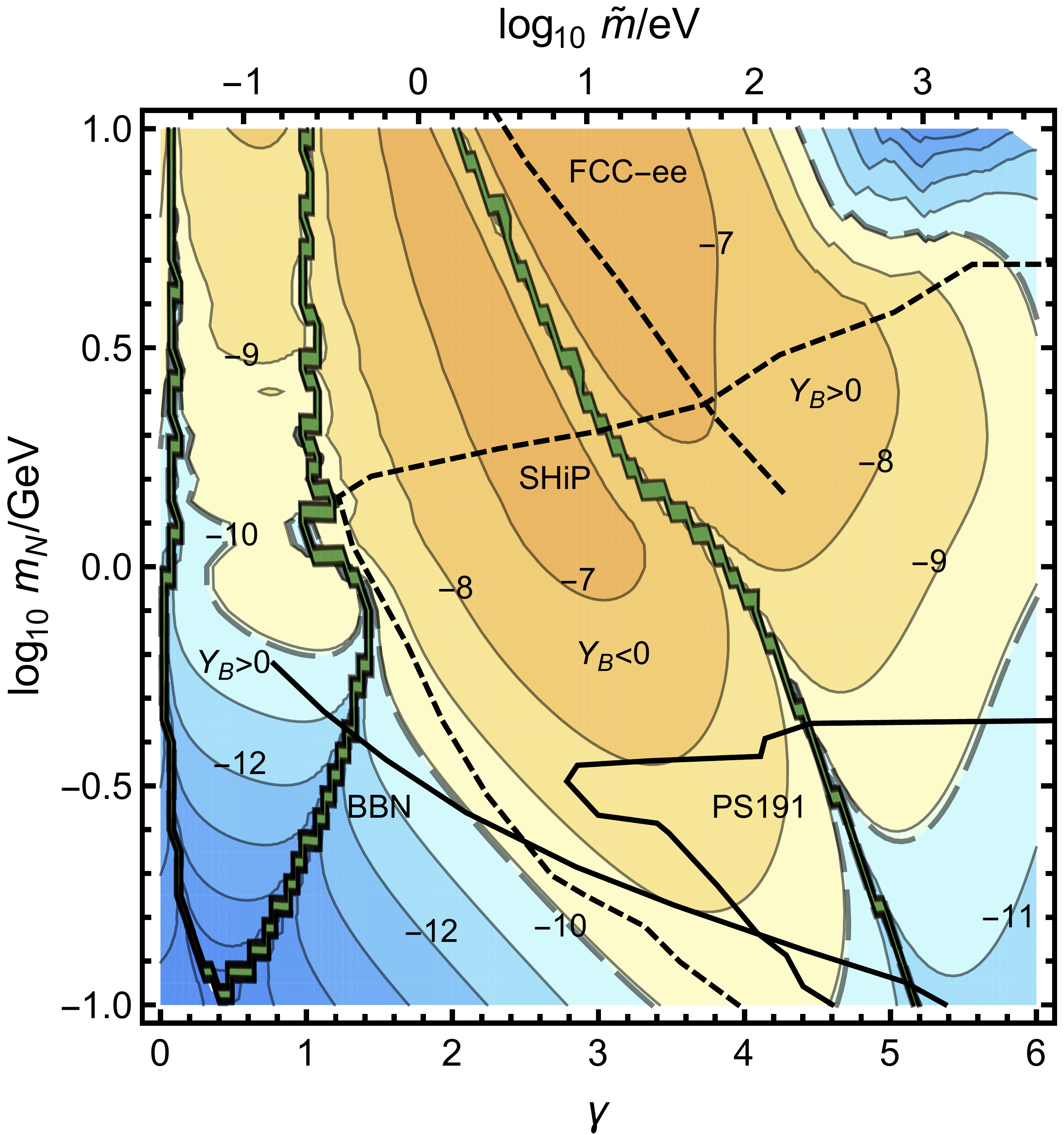}\qquad\qquad\qquad
\includegraphics[width=0.33\textwidth]{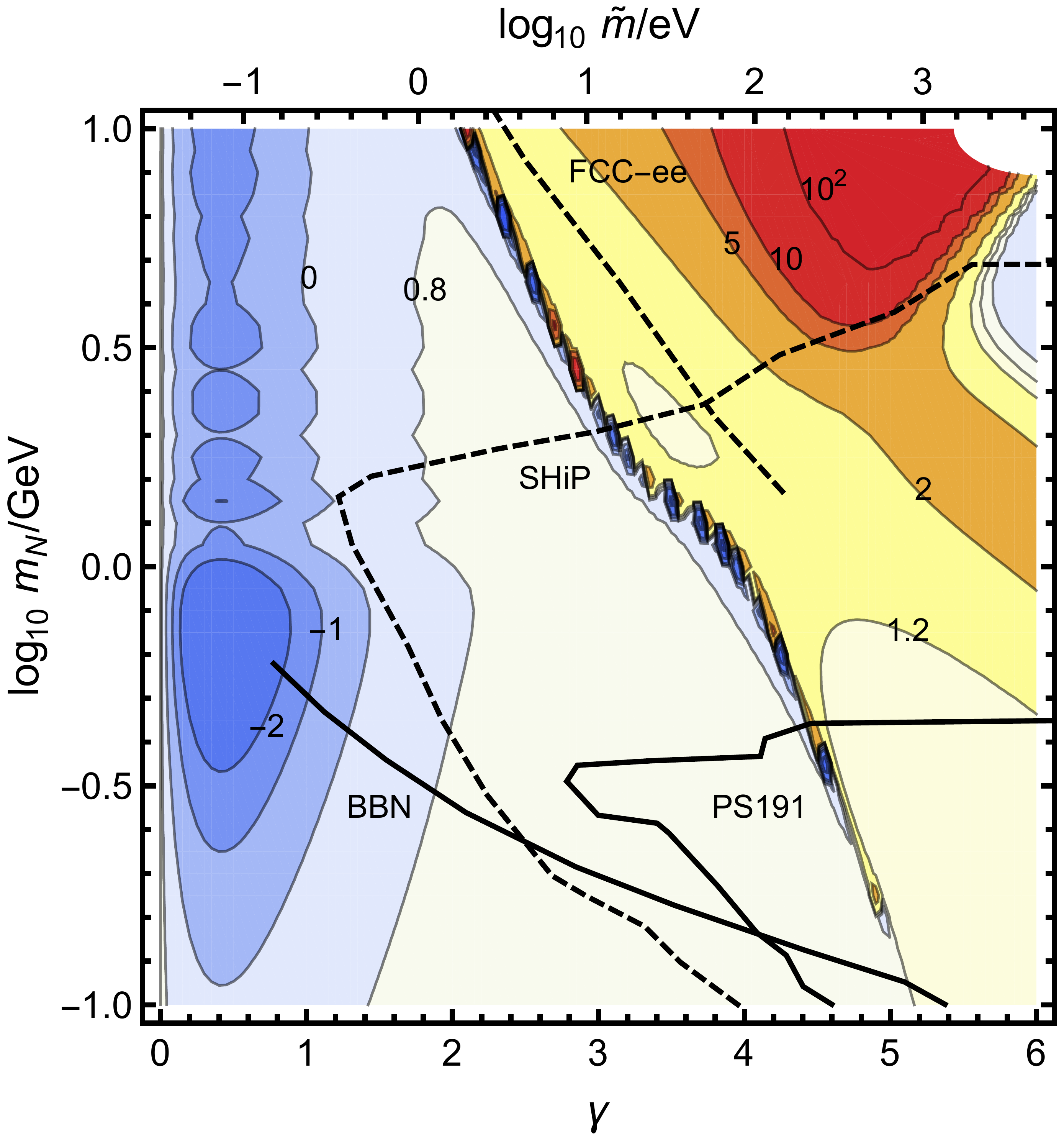}
\caption{Results for $\Delta m_N/m_N = 10^{-10}$, $\alpha_2=3 \pi/2$.\label{fig:10_D}}
\end{figure*}

\begin{figure*}[p]
\includegraphics[width=0.33\textwidth]{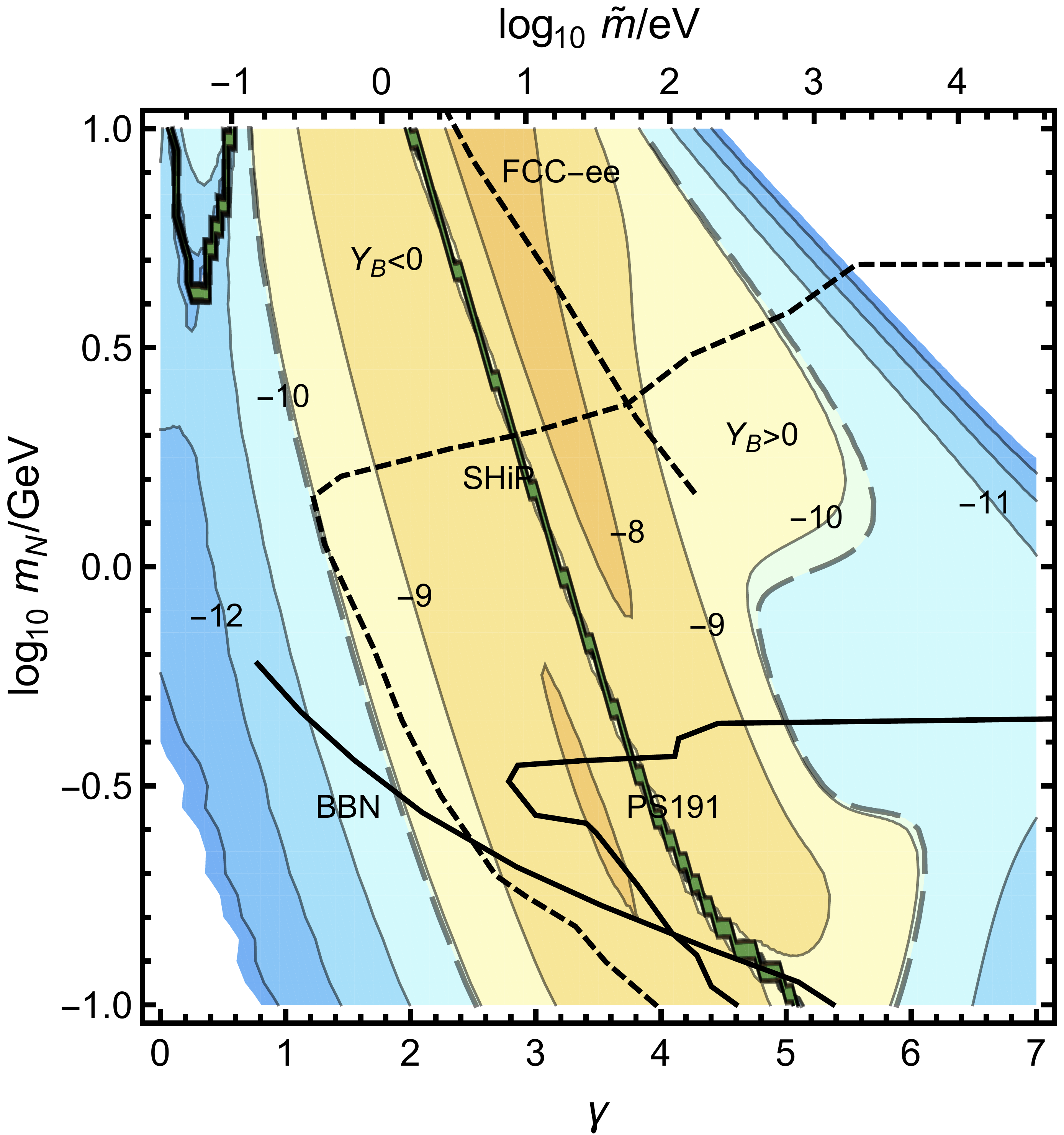}\qquad\qquad\qquad
\includegraphics[width=0.33\textwidth]{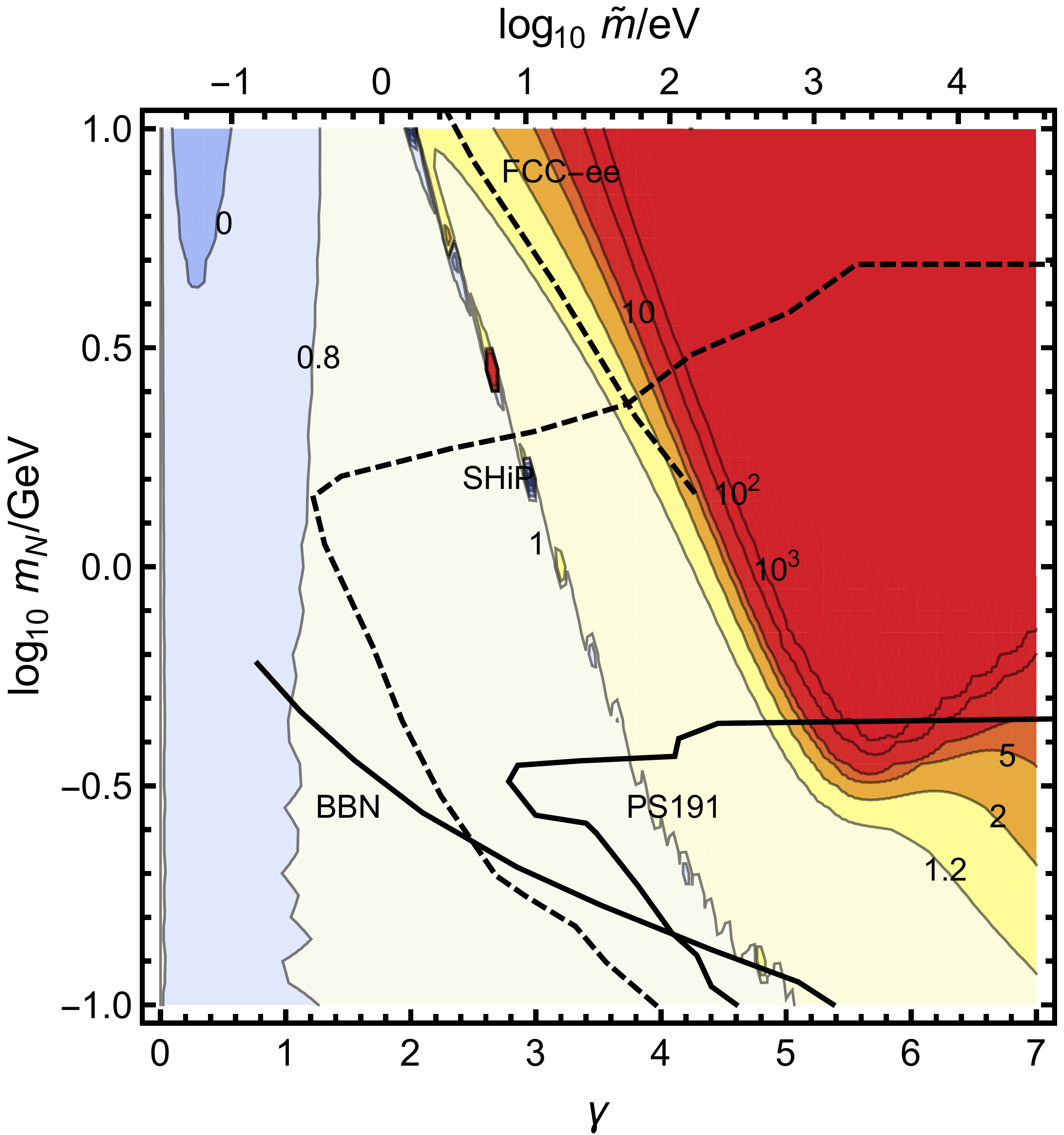}
\caption{Results for $\Delta m_N/m_N = 10^{-8}$, $\alpha_2=\pi/2$.\label{fig:8}}
\end{figure*}
\begin{figure*}[p]
\includegraphics[width=0.33\textwidth]{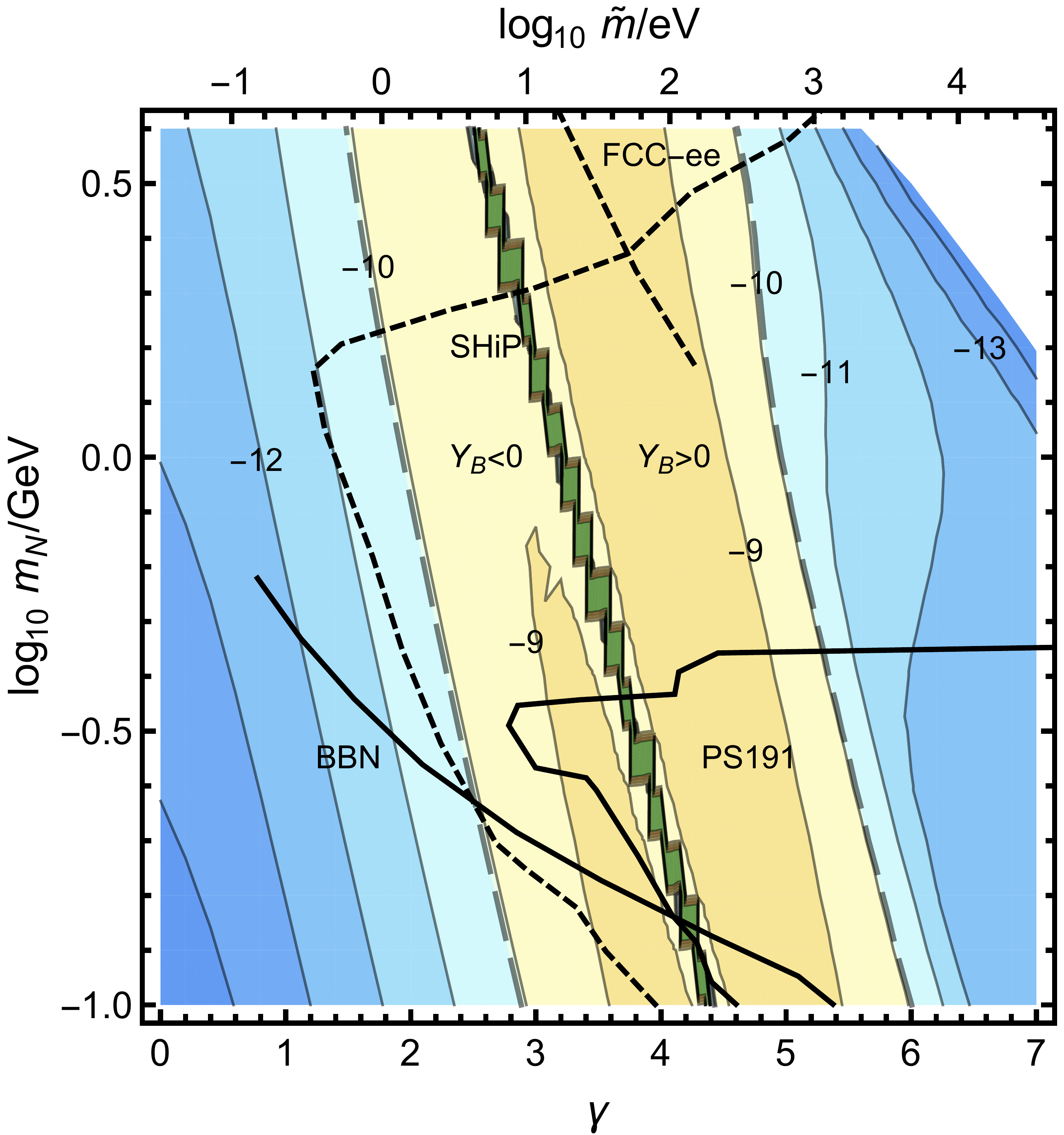}\qquad\qquad\qquad
\includegraphics[width=0.33\textwidth]{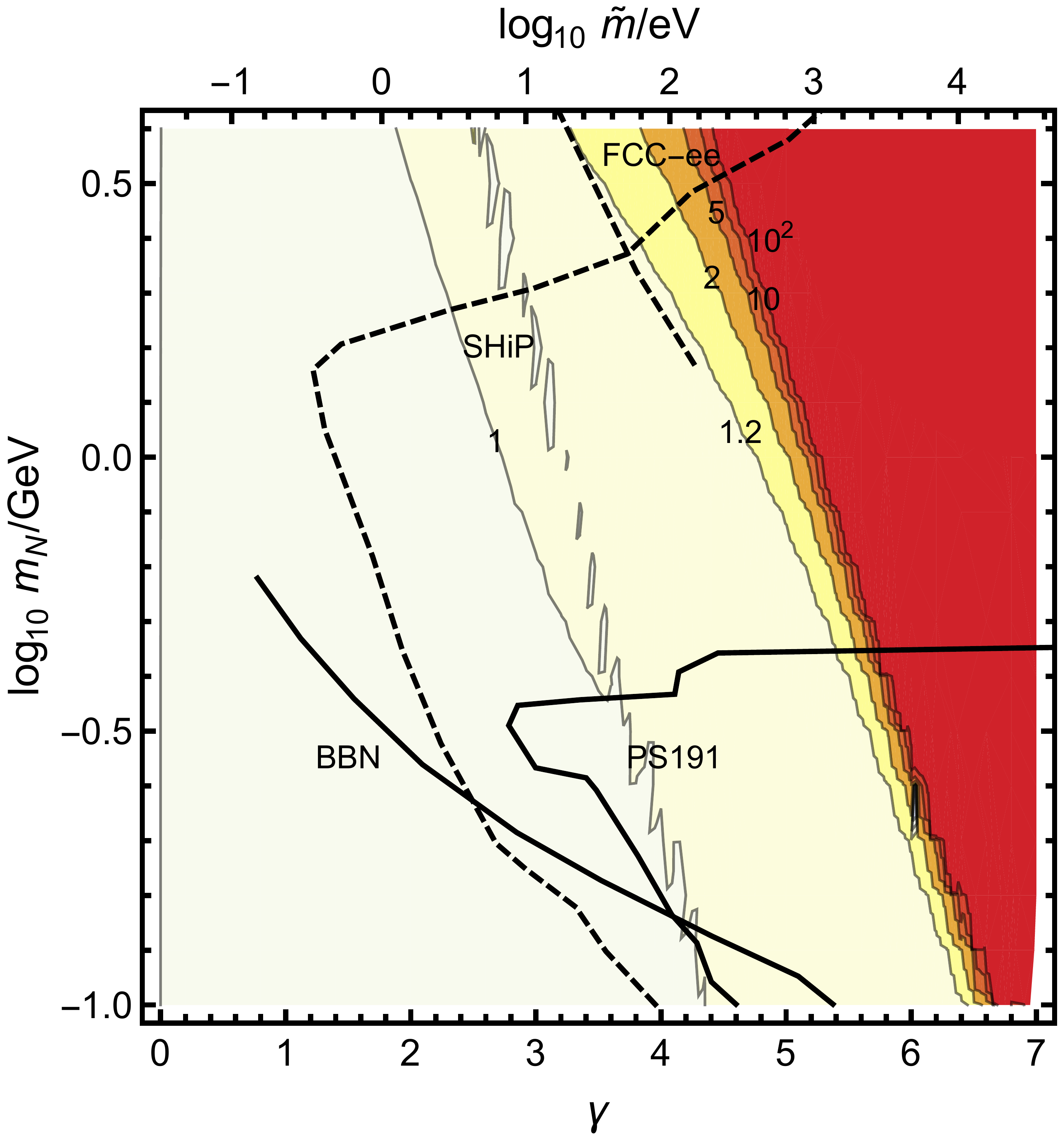}
\caption{Results for $\Delta m_N/m_N = 10^{-6}$, $\alpha_2=3 \pi/2$\label{fig:6}.}
\end{figure*}

We now consider the realistic situation with 3 SM lepton flavours, with seesaw parameters fulfilling all neutrino-mass constraints (with, here too, 2 RH neutrinos for simplicity). This allows us to identify the regions of parameter space where the LV Higgs-decay contribution is important as compared to the LC ARS one.

We fit the light-neutrino oscillation data by using the so-called Casas-Ibarra parametrization~\cite{Casas:2001sr}
\begin{equation}
h \ = \ - i \, \frac{\sqrt{2}}{v} \, U_{PMNS} \cdot \sqrt{m_\nu^{diag}} \cdot R \cdot m_N^{diag} \;,
\end{equation} 
where $v = \unit[246]{GeV}$, with $m_\nu^{diag}$ and $ m_N^{diag}$ denoting, in the mass eigenbasis, the light and heavy neutrino mass matrices,  respectively.  For the PMNS matrix we use the standard parametrization of~\cite{Olive:2016xmw}, fixing all parameters to their best-fit values~\cite{Olive:2016xmw} and $\delta = -\pi/2$. The Majorana phase $\alpha_1$ plays no role in what follows, unlike the $\alpha_2$ phase. Thus we will present our results for different values of $\alpha_2$. For definiteness we choose a normal hierarchy of masses. In the 2 RH-neutrino case considered here, the orthogonal matrix $R$ can be expressed in terms of a single complex angle $\omega = w + i \gamma$. We fix $w$ to the value which maximizes $\delta^{LV}$,  $w= \pi/4$. For $w$ fixed in this way the Casas-Ibarra parameter $\gamma$ 
has a one-to-one relation with the effective neutrino mass $\widetilde m \equiv v^2 (h^\dag h)_{11}/m_N$.

Starting here too from no RH neutrinos at $T>>>T_s$,  we have integrated numerically the evolution equations.
 The final $n_B/n_\gamma$ baryon asymmetry obtained is plotted as a function of $m_N$ and $\gamma$ in the left panels of Fig.~\ref{fig:10_A}--\ref{fig:6}, for various values of $\Delta m_N/m_N$ and $\alpha_2$\footnote{As already mentioned in Section \ref{sec:oneleptonflavour} for simplicity we have not included here the thermal masses for the RH neutrinos, whose proper treatment in the density-matrix formalism requires a $T$-dependent diagonalization of their mass matrix. However, their effect is expected to be important only for small mass splitting with quite low values of $m_N$.}. These plots also show the relevant existing constraints and future prospects of the proposed SHiP~\cite{Alekhin:2015byh} and FCC-ee~\cite{Antusch:2016vyf} experiments. We have also calculated the ARS lepton-number conserving contribution, obtained by setting to zero the LV rates in the evolution equations, as done previously in the existing literature. The results for the ratio of the full (LV Higgs decay + LC ARS) result to the ARS one are shown in the right panels of Fig.~\ref{fig:10_A}--\ref{fig:6}. The regions left in very light grey on these plots are the ones where this ratio lies between 0.8 and 1.2, i.e.~the regions where the incorporation of the LV part does not change sizeably the final result. 

Let us start with a mass splitting $\Delta m_N/m_N = 10^{-10}$. Fig.~\ref{fig:10_A}--\ref{fig:10_D} show the results for this mass splitting and $\alpha_2 = 0, \pi/2, \pi, 3\pi/2$. Although the precise quantitative values (and sign) of the ratio between the full and ARS results depend strongly on the phase $\alpha_2$, the overall picture is quite stable. For small $\gamma$, i.e.~for values of the Yukawa couplings of order of the ones given by the basic seesaw scaling, the two contributions have similar size, with the LV part dominating in a large region. This behaviour is already captured by the analytic solutions given in Sec.~\ref{sec:analytic}, and has been argued already in~\cite{Hambye:2016sby}. 

In addition, for large values of $\gamma$, the LV Higgs-decay contribution becomes generically dominant, even by many orders of magnitude\footnote{In the plots showing the total to ARS result ratio (right panels), we exhibit contours only up to $10^2$ or $10^3$ because for larger values the numerical noise due to the smallness of the ARS one, which appear in the denominator of the ratio, becomes important.} with respect to the ARS one. These large values of $\gamma$ (or equivalently $\widetilde m$) can be obtained generically if lepton number is a symmetry which is approximately conserved. Such a symmetry gives at  the same time a quasi-degenerate spectrum for 2 RH neutrinos, as necessary here. This large-$\gamma$ region covers a large fraction of the region of parameters accessible at SHiP and FCC-ee, i.e.~where these experiments could directly produce the RH neutrino(s). The origin of this LV dominance is clear:  for large $\gamma$ we are in the strong washout regime. In this regime, as already explained at the end of Section~\ref{sec:oneleptonflavour},
the $\gamma_W^{LC}$ washout part will effectively wash out the ARS LC part, whereas the $\gamma_W^{LV}$ washout part, which is suppressed by an extra $m_N^2/T^2$ factor (see \eqref{WQLCrate} and \eqref{WQLVrate}), will wash out less the LV part (and only at later times due to this factor), resulting in a dominant LV part.

In Fig.~\ref{fig:8} and~\ref{fig:6} we plot the results for $\Delta m_N/m_N$ equal to $10^{-8}$ and $10^{-6}$, respectively. As already emphasized in Sec.~\ref{sec:analytic}, for the weak-washout regime $\widetilde m \approx m_\nu$, the larger is $\Delta m_N/m_N$, the less is important the LV contribution. However, the LV effects are still important in the strong-washout region and still dominate by many orders of magnitude. Nevertheless, in this case the parameter space leading to successful leptogenesis shrinks, and the region of successful leptogenesis where the LV contribution dominates becomes smaller, even if still significant. For $\Delta m_N/m_N = 10^{-6}$, from the right panel of Fig.~\ref{fig:6} we see that the LV contribution is still largely dominant for $m_N \approx \unit[3]{GeV}$, $\gamma \approx 4-5$. For larger 
$m_N$ this dominance is expected to increase further but in this case the numerical solution of the density-matrix equations becomes computationally challenging and a more sophisticated numerical code would be needed (especially for even larger values of $\Delta m_N/m_N$).


\begin{figure}
\includegraphics[width=0.33\textwidth]{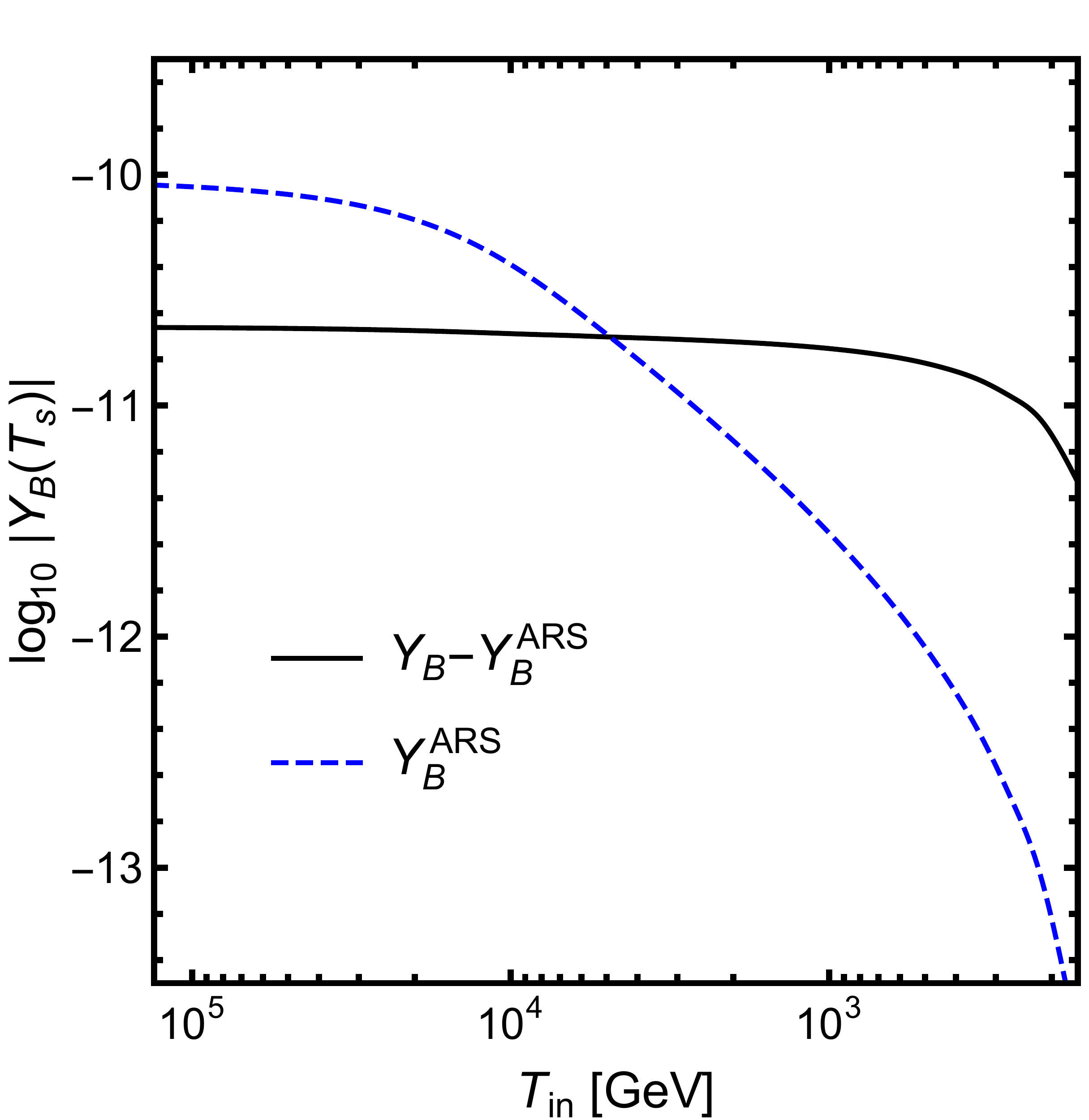}
\caption{The final $n_B/s$ asymmetry obtained as a function of the initial temperature $T_{in}$ (where one assumes no RH neutrinos to start with) for the LV Higgs-decay and ARS LC contributions. We have chosen: $m_N=\unit[10]{GeV}$, $\gamma=0.5$, $\Delta m_N/m_N = 10^{-7}$, $\alpha_2=0$.\label{fig:Tin}}
\end{figure}

We conclude this section by discussing the dependence on the initial ``reheating" temperature $T_{in}$, where one has assumed that there are no RH neutrinos to start with. As already argued in~\cite{Hambye:2016sby}, the LV Higgs-decay mechanism is a low-scale one, in the sense that the asymmetry is generated mainly at temperatures close to the sphaleron decoupling one. Thus, it does not rely on UV physics well above the electroweak scale. The LC flavoured asymmetries in the ARS scenario, instead, are mainly produced at $T \approx T_{osc} = (M_0 \Delta m_N^2)^{1/3}$ (at least in the weak-washout regime), which easily lies well above $T_s$. We illustrate this in Fig.~\ref{fig:Tin}, where we plot the LC and LV contributions as a function of the initial temperature $T_{in}$ in the weak-washout regime for $\Delta m_N/m_N = 10^{-7}$. The former is obtained, again, by setting to zero the LV rates, whereas the later is obtained as the total one minus the LC one, since in the weak-washout regime the two contributions are additive. For the parameters chosen in Fig.~\ref{fig:Tin}, even though the LV contribution is subdominant when $T_{in} > T_{osc}$, it becomes dominant, even by 2 orders of magnitude, if the initial temperature of the evolution is smaller.

\section{Conclusions}

In summary, we have computed, in the density-matrix formalism, the contribution to the baryon asymmetry of the total lepton-number violating Higgs-doublet decay and inverse decay. This contribution has been obtained by deriving the corresponding density-matrix evolution equations, and by integrating them. The results obtained confirm the ones derived in the CP-violating decay formalism~\cite{Hambye:2016sby}: this contribution is fully 
relevant for large parts of the parameter space. This is the case even if the L-violating processes are suppressed by a factor of $m^2_N/T^2$ with respect to the L-conserving ones, by virtue of a different parametric dependence, in particular in the Yukawa couplings and rates.

In the one lepton-flavour case the results for the LV contribution of both formalisms can be easily compared and show a nice qualitative agreement, or even quantitative agreement for not too large values of the Yukawa couplings. For larger Yukawa couplings, leading to strong washout, the density-matrix formalism gives a larger asymmetry because the washout of the LV contribution turns out to be much less efficient than what we could have expected at first sight.

In the realistic 3 lepton-flavour case, and for relatively small RH-neutrino mass splitting, the LV contribution dominates over the ARS LC contribution for Yukawa couplings of the order of the ones given by the usual neutrino-mass seesaw formula without cancellations, and also for  Yukawa couplings much larger than these. This latter dominance is, here too, due to the fact that the washout of the LV contribution is not easily efficient, and is actually much less efficient than the washout of the ARS LC contribution.
For larger mass splitting this dominance of the LV contribution remains generic for large Yukawa couplings. Also, as a result of the fact that the LV part involves the mass insertion $m_N^2$ factor, the larger $m_N$ the more the LV part dominates over the ARS LC part. Generic dominance of the LV contribution is also obtained for low reheating temperature. 

All the above takes into account exclusively the baryon asymmetry produced by the Higgs-doublet decay and inverse decay.
It is known that other processes can be fully relevant, in particular the top-quark and gauge scattering and the infrared-enhanced gauge corrections to Higgs decay \cite{Besak:2012qm,Ghiglieri:2016xye,Ghiglieri:2017gjz}.
These processes are expected to generate a baryon asymmetry in the same way because they involve as a subprocess the same Higgs-doublet transition into a RH neutrino and a SM lepton. One could therefore anticipate that, including them, the rates will be enhanced by a factor of 3 to 6 in the same way as for the ARS LC contribution~\cite{Besak:2012qm}, thus enhancing the asymmetry produced too.  
A fully quantitative calculation of the baryon asymmetry  produced would nevertheless require a proper inclusion of all these relevant LC and LV processes. 


\bigskip

\acknowledgments
We thank B. Garbrecht for useful comments on the washout issue.
This work is supported by the FNRS-FRS, the FRIA, the IISN, a ULB-ARC, the Belgian Science Policy (IAP VI-11) and a ULB Postdoctoral Fellowship.

\begin{appendix}

\section{Analytic solutions in the weak-washout and many-oscillations regime}\label{app:analytic}

In this Appendix we derive the analytic solutions in the weak-washout and many-oscillations regime, Eqs.~(\ref{LFanalytsol}) and (\ref{LVanalytsol}) . The expansion of the Universe is included in the Boltzmann equations~\eqref{eq:rate_eq}, \eqref{eq:rate_eq_bar} and \eqref{eq:rate_eq_asym} in which we replace the number densities $n$ by the corresponding yields
\begin{equation}
\rho \equiv \frac{n^N}{s} \;, \qquad \bar \rho \equiv \frac{\bar n^N}{s} \;, \qquad Y_l \equiv \frac{n^L_l}{s} \;, 
\end{equation}
where $s$ is the entropy density of the Universe, i.e.~we make the substitution $dn/dt \to s H(z) z dY/dz$, with $z \equiv T_s/T$. The sphaleron decoupling temperature is $T_s \simeq \unit[131.7]{GeV}$ and $H(z)$ is the Hubble parameter.
\subsection{ARS mechanism} 
In the weak-washout regime the asymmetry generated by LC processes (i.e.~by the ARS mechanism) can be obtained iteratively, following~\cite{Asaka:2005pn}, starting form $\rho(0) = \bar \rho(0) = Y_l(0) = 0$. By neglecting the LV rates, by absorbing the commutator terms of~\eqref{eq:rate_eq} and \eqref{eq:rate_eq_bar} in the other terms of these equations (by going to an ``interaction picture'' in the usual way) and by performing one iteration (i.e.~plugging $\rho(0) = \bar \rho(0) = 0$ on the RH side of the equations) we obtain
\begin{equation}
Y_l(z) \ \simeq \ 4 \, (\kappa^{LC})^2 \rho_{eq} \, (h^\dag h)_{11} (h^\dag h)_{22} \, \delta_l^{LC} \! \int_0^z \!\! \mathrm{Im} f(z') d z' \;,
\label{fint}
\end{equation}
with $\kappa^{X} \equiv 4 \pi^2/(3 \zeta(3)) (M_0/T_c) \, \alpha^X$,  $\rho_{eq} \simeq 135 \zeta(3)/(8 \pi^4 g_*)$ (and $g_*$ is the effective number of relativistic degrees of freedom). The $f$ function is 
\begin{equation}
f(z) \ \equiv \ \int_0^z \!\! d z' e^{ i \frac{\Delta m_N^2}{3 \mu^2_{osc}}  (z^3 - z'^3)} \;,
\end{equation}
with $\Delta m_N^2 \equiv m_{N,2}^2 - m_{N,1}^2$, $\mu^{2}_{osc} \equiv  2 T_c^3/(\langle T/|\ve k|  \rangle M_0) \simeq (\unit[3.75]{keV})^2$.
The quantity $\delta_l^{LC}$ is the usual SM lepton-number conserving CP violating Yukawa coupling combination
\begin{equation}
\delta^{LC}_l \ = \ \frac{\mathrm{Im}\big[h_{l1}^* h_{l2} (h^\dag h)_{21}\big]}{(h^\dag h)_{11} (h^\dag h)_{22}}
\end{equation}
In the regime $\Delta m_N^2 \gg \mu^2_{osc}$, $\mathrm{Im} f(z)$ performs many oscillations (of increasing frequency) before the sphalerons decouple, and its integral in (\ref{fint}) saturates at early times to the value
\begin{equation}
 \mathrm{Im} \int_0^{z} \!\! f(z') d z' \ \approx \ 1.4 \times \bigg(\frac{\mu^2_{osc}}{\Delta m^2_N}\bigg)^{\frac{2}{3}} \;.
\end{equation}
Therefore, at $O(h^4)$, after the first few oscillations (at $z \approx z_{osc} \equiv T_c/(M_0 \Delta m_N^2)^{1/3}$) one obtains
\begin{equation}
Y_l(z) \ \simeq \ 4 \times 1.4 \,(\kappa^{LC})^2 \rho_{eq} \, (h^\dag h)_{11} (h^\dag h)_{22} \, \delta_l^{LC}\bigg(\frac{\mu^2_{osc}}{\Delta m^2_N}\bigg)^{\frac{2}{3}} \!.
\end{equation}
Notice that, since $\sum_l \delta^{LC}_l = 0$, the total SM-lepton asymmetry vanishes at $O(h^4)$,
\begin{equation}
\sum_l Y_l =  0 \quad \text{at} \quad O(h^4) \;.
\end{equation}
In order to obtain a non-vanishing total asymmetry one has to perform one iteration more in the solution of the Boltzmann equations, in particular taking into account at leading order the effect of a flavour-asymmetric washout. By exploiting the fact that, neglecting lepton-number violating processes, $Y_{LC} \equiv \sum_l Y_l= \tr(\rho - \bar \rho)$, one finds at $O(h^6)$:
\begin{align}
&Y_{LC}(z=1) \ \simeq \notag\\ \ &- \, 2 \times 1.4 \,(\kappa^{LC})^2 \, \kappa^{LC}_W \rho_{eq} \,  \bigg(\frac{\mu^2_{osc}}{\Delta m^2_N}\bigg)^{\frac{2}{3}} \, (h^\dag h)_{11} (h^\dag h)_{22} \, \notag\\
&\times \sum_l \delta_l^{LC} (h h^\dag)_{ll} 
\end{align}
which gives approximately Eq.~(\ref{LFanalytsol}).

\subsection{L-violating Higgs-decay contribution}
In the same regime as before we can now calculate analytically the contribution of the lepton-number violating processes. Proceeding iteratively, there are two contributions: the first one comes from $\gamma^{LV}$ in the equations for $\rho$ and $\bar \rho$ and $\gamma^{LC}$ in the equation for $Y_l$. We find this contribution to be
\begin{align}
Y_{LV}(z) \ &\supset \ 4 \, \kappa^{LC} \kappa^{LV} \frac{m_N^2}{T_c^2} \rho_{eq} \, (h^\dag h)_{11} (h^\dag h)_{22} \, \delta^{LV} \notag\\
&\times \, \int_0^z \!\! \mathrm{Im}\,g(z') \,d z' \;,
\end{align}
with
\begin{equation}
g(z) \ \equiv \ \int_0^z \!\! d z'\, z'^2 \, e^{i \frac{\Delta m_N^2}{3 \mu^2_{osc}}  (z^3 - z'^3)} \;,
\end{equation}
and where $\delta_l^{LV}$ is the total lepton-number violating and CP-violating combination of the Yukawa couplings
\begin{equation}
\delta^{LV}_l \ = \ \frac{\mathrm{Im}\big[h_{l1}^* h_{l2} (h^\dag h)_{12}\big]}{(h^\dag h)_{11} (h^\dag h)_{22}}\;.
\end{equation}
Notice that we have already summed over SM-lepton flavours, since $\delta^{LV} \equiv \sum_l \delta^{LV}_l \neq 0$, i.e.~the LV Higgs-decay mechanism gives rise to an unflavoured asymmetry already at $O(h^4)$, differently from the ARS one.
In the regime $\Delta m_N^2 \gg \mu^2_{osc}$ the integral of $\mathrm{Im}\,g(z')$ approaches the behaviour 
\begin{equation}
 \mathrm{Im} \int_0^{z} \!\! g(z') \, d z' \ \approx \ z \, \frac{\mu^2_{osc}}{\Delta m^2_N} \;.
\end{equation}
up to a residual oscillating term (which we neglect) whose frequency increases and amplitude decreases with time. Thus,
\begin{align}\label{eq:first_contrib}
Y_{LV}(z=1) \ &\supset \ 4 \, \kappa^{LC} \kappa^{LV} \frac{m_N^2}{T_c^2} \rho_{eq} \, (h^\dag h)_{11} (h^\dag h)_{22} \, \delta^{LV} \notag\\
& \times \, \frac{\mu^2_{osc}}{\Delta m^2_N} \;,
\end{align}

The second contribution comes from the iteration with $\gamma^{LC}$ in the equations for $\rho$ and $\bar \rho$ and $\gamma^{LV}$ in the equation for $\mu$. This is given by
\begin{align}
Y_{LV}(z) \ &\supset \ 4 \, \kappa^{LC} \kappa^{LV} \frac{m_N^2}{T_c^2} \rho_{eq} \, (h^\dag h)_{11} (h^\dag h)_{22} \, \delta^{LV} \notag\\
&\times \, \int_0^z \!\! z'^2 \, \mathrm{Im}\,f(z') \, d z' \;,
\end{align}
The integral is again found to approach the behaviour 
\begin{equation}
 \mathrm{Im} \int_0^{z} \!\! z'^2 \, f(z') d z' \ \approx \ z \, \frac{\mu^2_{osc}}{\Delta m^2_N} \;,
\end{equation}
up to an oscillating term which with time increases its frequency but does not change its amplitude (equal to $(\mu^2_{osc}/\Delta m_N^2)^{4/3}$). Neglecting this term we find that this second contribution is the same as the first one \eqref{eq:first_contrib}. The sum of both terms leads to Eq.~(\ref{LVanalytsol}).
\end{appendix}

\end{document}